\documentclass[fleqn,usenatbib]{mnras}

\usepackage[T1]{fontenc}

\DeclareRobustCommand{\VAN}[3]{#2}
\let\VANthebibliography\thebibliography
\def\thebibliography{\DeclareRobustCommand{\VAN}[3]{##3}\VANthebibliography}

\usepackage{graphicx}	
\usepackage{amsmath}
\usepackage{physics}
\usepackage{xurl}
\usepackage{comment}
\usepackage{multirow}
\usepackage{newtxtext,newtxmath}

\newcommand{\e}[1]{\ensuremath{\times 10^{#1}}}
\newcommand{\Msun}[1]{\ensuremath{\textup{M}_\odot}}
\newcommand{\refeq}[1]{Eq.~\eqref{#1}}
\newcommand{\reffig}[1]{Fig.~\ref{#1}}
\newcommand{\reftab}[1]{Table~\ref{#1}}

\newcommand{\hmpc}{\,h^{-1}\,{\rm Mpc}}

\newcommand{\mycmt}[1]{{{#1}}}

\title[3-parameter LRG SHAM]
{Model BOSS \& eBOSS Luminous Red Galaxies at $0.2 < z < 1.0$ using SubHalo Abundance Matching with 3 parameters}
\author[Jiaxi Yu et al.]
{\parbox[t]{\textwidth}{\vspace{-0.8cm}Jiaxi Yu,$^{1}$\thanks{E-mail: jiaxi.yu@epfl.ch}
Cheng Zhao,$^{1}$\thanks{E-mail: cheng.zhao@epfl.ch}
Chia-Hsun Chuang,$^{2,3}$\thanks{E-mail:albert.chuang@utah.edu}
Julian E. Bautista, $^{4}$ 
Ginevra Favole,$^{1}$
Jean-Paul Kneib,$^{1,5}$ 
Faizan G. Mohammad, $^{6}$
Ashley J. Ross, $^{7}$
Anand Raichoor, $^{8}$
Charling Tao, $^{4}$ 
Kyle Dawson, $^{2}$
Graziano Rossi $^{9}$}
\vspace*{15pt}\\
$^{1}$Laboratory of Astrophysics, \'Ecole Polytechnique F\'ed\'erale de Lausanne (EPFL), Observatoire de Sauverny, CH-1290 Versoix, Switzerland\\
$^{2}$Department of Physics and Astronomy, University of Utah, Salt Lake City, UT 84112, USA\\
$^{3}$Kavli Institute for Particle Astrophysics and Cosmology, Stanford University, 452 Lomita Mall, Stanford, CA 94305, USA\\
$^{4}$Aix Marseille Univ, CNRS/IN2P3, CPPM, Marseille, France \\ 
$^{5}$Aix Marseille Universit\'e, CNRS, LAM (Laboratoire d'Astrophysique de Marseille) UMR 7326, F13388, Marseille, France\\
$^{6}$Waterloo Centre for Astrophysics, Dept. of Physics and Astronomy, University of Waterloo, 200 University Ave. W., Waterloo ON N2L 3G1, Canada\\
$^{7}$Department of Astronomy, The Ohio State University, 140 W. 18th Ave., Columbus, OH 43210, USA\\
$^{8}$Lawrence Berkeley National Laboratory, 1 Cyclotron Road, Berkeley, CA 94720, USA\\
$^{9}$Department of Astronomy and Space Science, Sejong University, 209, Neungdong-ro, Gwangjin-gu, Seoul, South Korea
\vspace{-1.2cm}\\
}

\date{Accepted XXX. Received YYY; in original form ZZZ}

\pubyear{}

\begin{document}
\label{firstpage}
\pagerange{\pageref{firstpage}--\pageref{lastpage}}
\maketitle

\begin{abstract}
SubHalo Abundance Matching (SHAM) is an empirical method for constructing galaxy catalogues based on high-resolution $N$-body simulations. We apply SHAM on the UNIT simulation to simulate SDSS BOSS/eBOSS Luminous Red Galaxies (LRGs) within a wide redshift range of $0.2 < z < 1.0$. Besides the typical SHAM scatter parameter $\sigma$, we include $v_{\rm smear}$ and $V_{\rm ceil}$ to take into account the redshift uncertainty and the galaxy incompleteness respectively. These two additional parameters are critical for reproducing the observed 2PCF multipoles on 5--25$\hmpc$. The redshift uncertainties obtained from the best-fitting $v_{\rm smear}$ agree with those measured from repeat observations for all SDSS LRGs except for the LOWZ sample. We explore several potential systematics but none of them can explain the discrepancy found in LOWZ. Our explanation is that the LOWZ galaxies might contain another type of galaxies which needs to be treated differently. The evolution of the measured $\sigma$ and $V_{\rm ceil}$ also reveals that the incompleteness of eBOSS galaxies decreases with the redshift. This is the consequence of the magnitude lower limit applied in eBOSS LRG target selection. Our SHAM also set upper limits for the intrinsic scatter of the galaxy--halo relation given a complete galaxy sample: $\sigma_{\rm int}<0.31$ for LOWZ at $0.2<z<0.33$, $\sigma_{\rm int}<0.36$ for LOWZ at $0.33<z<0.43$, and $\sigma_{\rm int}<0.46$ for CMASS at $0.43<z<0.51$. The projected 2PCFs of our SHAM galaxies also agree with the observational ones on the 2PCF fitting range.
\end{abstract}



\section{Introduction}
Lambda-Cold-Dark-Matter ($\Lambda$CDM) has been the standard cosmological model since the 1990s. Under this framework, observations confirm that dark matter (DM) is the dominant matter component of the Universe \citep[e.g.,][]{cmb2018}. DM particles interact only through gravity and their gravitational evolution is assumed to start from a primordial Gaussian random field with perturbations. If the over-density arising from the perturbation on small scales is large enough, the infall of matter can overcome the expansion of Universe, and form gravitational-bound clusters, i.e., DM haloes. Perturbations on large scales evolve into web-like structures known as the cosmic web \citep{cosmicweb1996}. Baryonic matter may be captured by haloes with deep potential wells and further evolve into galaxies \citep{press_schechter1974, white1978}, a good tracer of the invisible cosmic web.  

Galaxy surveys explore the large-scale structure of the Universe by measuring the redshifts of millions of galaxies and quasars. Their positions can be used to calculate the two-point correlation function (2PCF) that encodes the history of the universe expansion and the structure growth. The Baryon Oscillation Spectroscopic Survey \citep[BOSS, 2008--2014;][]{Dawson2013} is the largest project in the third-stage Sloan Digital Spectroscopic Survey\footnote{\url{http://www.sdss.org/}} \citep[SDSS-III;][]{SDSSIII}. It has collected the spectra of more than 1.5 million Luminous Red Galaxies \citep[LRGs;][]{reid_sdss-iii_2016}, the brightest red galaxies in the Universe. Its extended version, eBOSS \citep[2014--2020;][]{Dawson2015} in SDSS-IV \citep{SDSSIV} has probed another 300,000 LRGs \citep{ross_completed_2020}. Additionally, eBOSS has observed around 270,000 Emission Line Galaxies \citep[ELGs;][]{Anand2020}. They are bluer, star-forming galaxies that are abundant in the redshift range of $0.5< z< 2$, where there are active star formation processes \citep{Madau1998}. With such a big amount of tracers, both projects have achieved a percent-level precision in cosmological parameter measurements \citep{bossshadab, eBOSSshadab}. Ongoing surveys like the Dark Energy Spectroscopic Instrument \citep[DESI, 2020-2025;][]{DESICollaboration2016} is expected to provide tracers in a larger footprint and with higher-resolution spectra, and reach a higher precision in scientific results.

Meanwhile, $N$-body simulations can solve numerically the gravitational evolution equation of DM in $\Lambda$CDM. They are able to accurately describe the DM distribution in the non-linear regime down to individual haloes in a large volume and provide halo properties at any cosmic epoch. Thus $N$-body simulations help validate cosmological theories that are based on perturbation theories across all the scales, and makes it possible to compare theoretical predictions with observations. 

But there is a gap between dark-matter-only $N$-body simulations and observations. Since baryonic matter interacts with each other through all fundamental forces, their clustering properties do not necessarily follow those of DM on all scales \citep{white1978}. Moreover, the spatial distributions of different types of galaxies, such as LRGs and ELGs, can also be different in the cosmic web \citep[e.g.,][]{malavasi_vimos_2017, kraljic_galaxy_2018}. To bridge the gap, a galaxy--halo relation that modulates the clustering of haloes to match the observation is of great importance. Because it enables the direct comparison between the theory and the observation via simulation \citep{wechsler_connection_2018}. 

SubHalo Abundance Matching is a simple and intuitive empirical algorithm to model the galaxy--halo relation. As indicated by the name, SHAM makes use of DM haloes and their substructures that are dubbed subhaloes. The basic assumption of SHAM is a monotonic relation (not necessarily linear) between the galaxy luminosity (or stellar mass) and the halo mass \citep[or the circular velocity;][]{Kravtsov2004, vale2004, nagai2005, Conroy2006, Behroozi2010}. The algorithm applies the rank-ordering for both haloes and galaxies based on mass-related properties as just mentioned, and assigns galaxies to haloes/subhaloes one by one from the most massive end, until the SHAM galaxy catalogue reaches a desired number density. 

Galaxies from this mock catalogue follow the observed or reconstructed luminosity function \citep[e.g.,][]{tasitsiomi_modeling_2004, Conroy2006} or stellar mass function \citep[e.g.,][]{Guo2010, Rodriguez-Torres2016} by construction. Their two-point statistics are tuned to be consistent with the observed one by introducing a Gaussian scatter between the galaxy luminosity and the halo circular velocity \citep{tasitsiomi_modeling_2004} to include the scatter of the Tully--Fisher relation \citep{willick_homogeneous_1997,steinmetz_cosmological_1999}. Using the peak maximum circular velocity throughout accumulation history ($V_{\rm peak}$) instead of the halo mass further improves the performance of SHAM, since the galaxy accretion is free from the mass-stripping effect of subhaloes after reaching $V_{\rm peak}$ \citep{trujillo-gomez_galaxies_2011}. With all the improvements, SHAM catalogues of \citet{Rodriguez-Torres2016} predict the galaxy--halo relation that agrees with the weak-lensing observation from \citet{Shan2017}. Other work on SHAM has also included the effect of halo assembly bias and galaxy formation \citep[e.g.,][]{SHAM_assembly_bias_2016, SHAM_hydro_2021}. Due to the single-parameter feature and its good agreement on data, SHAM has become a useful tool in cosmological studies. 

While most of the SHAM studies choose to fit the projected 2PCF which marginalizes the effect along the line of sight, 2PCF multipoles can provide extra information in that direction. Due to the Redshift-Space Distortion (RSD), 2PCF quadrupole in the redshift space is vulnerable to the bias induced by the redshift uncertainty. This uncertainty mainly comes from the LRG redshift determination pipeline that uses broad absorption lines to determine the redshift \citep{bolton_spectral_2012}. For eBOSS LRG, \citet{ross_completed_2020} find the uncertainty measured by the repeat observations can be fit by a Gaussian function with a dispersion of 91.8 ${\rm km}\,{\rm s}^{-1}$. This is negligible for large-scale clustering analysis like Baryon Acoustic Oscillation (BAO) at over $80\hmpc$. Meanwhile, \citet{smith_completed_2020} demonstrates that the redshift uncertainty of eBOSS quasars influences the quadrupole on $r<60\hmpc$, affecting SHAM that mainly employs 2PCF on $r<40\hmpc$.

Moreover, the standard SHAM algorithm described above works well for bright galaxies with complete samples, as it matches halo masses and galaxy masses starting from the most massive ones. When the sample is incomplete due to the survey requirement (e.g. exclude targets with higher luminosity) or when the tracer is absent in massive haloes due to the quenching process \citep[e.g.,][]{kauffmann_environmental_2004, Dekel2006}, this implementation can be problematic \citep{favole_clustering_2016,Rodriguez-Torres2016}.

In this paper, we present a general 3-parameter SHAM algorithm for LRGs on 5-25$\hmpc$ that considers the redshift uncertainty effects and the galaxy completeness. In Section~\ref{data describe}, we describe the observational data, the $N$-body simulation and the galaxy mocks for covariance matrices. The calculation of 2PCFs and projected 2PCFs, the redshift uncertainty measurements, and the implementation of SHAM are introduced in Section~\ref{methods}. Section~\ref{Results} illustrates the good performance of SHAM in fitting the 2PCF, reproducing the redshift uncertainty, the galaxy incompleteness evolution and the projected 2PCF of the observation. Finally, we summarize our findings in Section~\ref{conclusion}. In our study, we use a flat $\Lambda$CDM cosmology with $\Omega_{\rm m}$ = 0.31 and $H_0=67.7\,{\rm km}\,{\rm s}^{-1}\,{\rm Mpc}^{-1}$, and the line-of-sight direction of the comoving space is along the $Z$-axis.

The BOSS \& eBOSS SHAM study is part of the final release of the eBOSS measurement. \citet{ross_completed_2020,lyke20a} describe in detail catalogues for the large-scale structure analysis. The mock challenge tasks are completed by \citet{alam20,avila20,rossi20a,smith_completed_2020} for systematics estimations. Galaxy mocks for covariance matrices used for cosmological analysis are constructed by \citet{lin20a,Zhao2020}. The BAO and RSD measurements\footnote{\url{https://www.sdss.org/science/final-bao-and-rsd-measurements}} are based on the clustering of LRGs at $0.6<z<1.0$ \citep{LRG_corr,gil-marin20a}, ELGs at $0.6<z<1.1$ \citep{Anand2020,tamone20a,demattia20a}, quasars (QSOs) at $0.8<z<2.2$ \citep{hou20a,neveux20a} and also the Ly$\alpha$ forest at $z>2.1$ \citep{2020duMasdesBourbouxH}. Their cosmological interpretations\footnote{see \url{https://www.sdss.org/science/cosmology-results-from-eboss}, and  \url{https://svn.sdss.org/public/data/eboss/DR16cosmo/tags/v1_0_1/}} in combination with the final BOSS results and other probes can be found in \citet{eBOSS_Cosmology}. eBOSS data also permit the analysis with cosmic voids \citep{Aubert20} and multiple tracers \citep{Wang20,ZhaoGB20,zhao2022}.

\section{Data}
\label{data describe}
\subsection{SDSS Galaxies}
We use the galaxy catalogues from the final data releases of SDSS-III BOSS \citep[DR12;][]{DR12} and SDSS-IV eBOSS \citep[DR16][]{DR16}. BOSS LRGs are composed of two sub-samples: LOWZ at $0.15<z<0.5$ and CMASS at $0.4 < z < 0.7$ \citep{reid_sdss-iii_2016}. For eBOSS, its LRG sample has redshift at $0.6<z<1.0$ \citep{ross_completed_2020}. Galaxy samples in the catalogues are pre-selected using the photometric information to ensure the observed samples are clean and abundant at a given redshift range for scientific targets. Their spectra are observed by fibers \citep{York2000,Gunn2006,Smee2013} and processed by the spectroscopic pipeline in order to determine their galaxy type and redshifts \citep{SDSSI/II2011,bolton_spectral_2012}. The catalogues used for clustering analysis (clustering catalogue hereafter) are composed of the 3D position of tracers, as well as several weights to eliminate systematics effects on 2PCF measurements. 

\subsubsection{Target Selection}
\label{target selection}
Target selection using the photometric information includes the signal-to-noise ratio (SNR) selection, the colour selection, the flux limits cut, and the star exclusion \citep{reid_sdss-iii_2016,prakash_sdss-iv_2016}. In general, those criteria are meant to have high quality LRG spectra, obtain the designed redshift range and number density, exclude the low-redshift, blue galaxies and ensure high-redshift successful rate, and remove stars from LRG samples, respectively. LOWZ at $0.2<z<0.4$ and CMASS $z<0.6$ are expected to be nearly volume-limited, i.e., they are complete \citep{reid_sdss-iii_2016}. The study of \citet{leauthaud_stripe_2016} confirms the LOWZ completeness but also finds CMASS is not as complete as \citet{reid_sdss-iii_2016} describes. 

For eBOSS LRG, there is a special magnitude cut, namely
\begin{equation}
    i \ge 19.9 ,
    \label{iband}
\end{equation}
which is set to avoid the BOSS CMASS galaxies \citep{prakash_sdss-iv_2016}. This lower limit corresponds roughly to an upper limit of the stellar mass, meaning that the most massive LRGs may have been excluded from the eBOSS samples. We use $V_{\rm ceil}$ to account for it and we will explain it in detail in Section~\ref{sham implementation}.

\subsubsection{Repeat Observations}
Repeat observations from SDSS aim at testing the reproducibility of the spectral measurements and obtaining a higher SNR by coadding multiple spectra \citep{Dawson2013}. We use them to estimate the redshift uncertainty statistically in Section~\ref{deltav calculation}. The redshifts of repeated samples from BOSS are determined by \textsc{idlspec2d} \citep{bolton_spectral_2012} and those from eBOSS LRGs are determined by \textsc{redrock} \footnote{\url{ https://github.com/desihub/redrock}}. 
Less than 5\% of galaxies in the clustering catalogues are observed twice, but those galaxies are good representatives of the clustering galaxies (Section~\ref{vsmear vs repeat}). 

\subsubsection{Galaxy Weights}
\label{galaxy weights}
The 2PCF contains the information of cosmological parameters and structure growth in its amplitude. But the amplitude can be easily modified by various systematics, e.g., photometric systematics, redshift failures, fibre collision effects and the varying galaxy number density at different redshift \citep{reid_sdss-iii_2016,ross_completed_2020}. The corresponding weights to remove their impacts are: $w_{\rm photosys}$ for all the photometric effects, $w_{\rm noz}$ for the redshift failure rate, $w_{\rm CP}$ for the missing close galaxy pairs due to the fibre collision and $w_{\rm FKP}$ for minimizing the cosmic variance. 

The fibre collision problem arises from the physical size of fibres (62$''$) \mycmt{that defines the minimum separation between two fibres. Hence galaxies with small angular distances} (i.e., close pairs) cannot be observed in a single exposure. \mycmt{For the BOSS data, the close-pair weight
\begin{equation}
    w_{\rm CP}=\frac{N_{\rm CP}+N_{\rm good}}{N_{\rm good}},
\end{equation}
is applied to the nearest tracer of collided targets to nullify the fibre collision effects. Here $N_{\rm CP}$ is the number of the half-missing tracer pairs and $N_{\rm good}$ is the number of tracers with good redshifts \citep{ross_completed_2020}. This nearest-neighbour approach is not able to fully correct the fibre collision effect \citep{Guo2012,Rodriguez-Torres2016}. Nevertheless, our SHAM fitting results are not biased by this effect (see Appendix~\ref{10-25 no bias}).}

eBOSS uses a different weighting scheme. The pairwise-inverse-probability (PIP) weighting proposed by \citet{bianchi17} is a better way to account for galaxy pairs with one missing in a single observation and the angular up-weighting (ANG) scheme introduced by \citet{percival17} can recover missing galaxy pairs with a distance smaller than the size of a fibre. \citet{Mohammad_2020} apply both PIP and ANG weights on eBOSS samples and provide unbiased galaxy clustering down to 0.1$\hmpc$.

The FKP weight is obtained by \citep{FKP1994}
\begin{equation}
    w_{\rm FKP} = \frac{1}{1+\overline{n}(z)P_0},
\end{equation}
where $\overline{n}(z)$ is the number density at redshift $z$ and $P_0$ is the amplitude of the observed power spectrum at $k\approx {0.15}\, h\,{\rm Mpc}^{-1}$. For LRGs in BOSS/eBOSS, we take $P_0={10000}\, h^{-3}\,{\rm Mpc}^{3}$ \citep{reid_sdss-iii_2016,ross_completed_2020}.

\subsection{\texorpdfstring{$N$}{N}-body Simulation: UNIT}
Future galaxy surveys will span to larger cosmological volumes with a deeper photometry. The resolution and effective volume of the $N$-body simulation should also keep up with the improvements. In our study, we use the Universe $N$-body simulations for the Investigation of Theoretical models from galaxy surveys\footnote{\url{http://www.unitsims.org/}} \citep[UNIT;][]{2019MNRAS.487...48C}. This is a high-resolution and large-effective-volume $N$-body simulation that uses the fixed-amplitude method to suppress the cosmic variance, thus increasing the effective volume \citep{Angulo2016}. 
Its effective volume is over 10 times larger than that of BOSS/eBOSS LRGs. UNIT has $4096^3$ particles in 1$\,h^{-3}\,{\rm Gpc}^{3}$ cubic box, and a mass resolution of 1.2\e{9}\,\Msun{}\,$h^{-1}$. Its cosmological parameters are $\Omega_m = 0.3089,~h \equiv H_0/100\,{\rm km}\,{\rm s}^{-1}\,{\rm Mpc}^{-1} = 0.6774,~ n_s = 0.9667,~ \sigma_8 = 0.8147$. The simulation evolves from $a(t)=0.01~(z=99)$ to $a(t)=1 ~(z=0)$, and produces 128 snapshots at different redshifts. \mycmt{In our study, we use 12 snapshots from $z=0.2760$ to $z=0.9011$ as shown in \reftab{snapshot}, and we take one simulation box in each snapshot given the large effective volume of UNIT (see Section~\ref{result fitting}).}

\subsection{Galaxy Mocks for Covariance Matrices}
\label{mocks}
We use galaxy mocks to calculate covariance matrices in Section~\ref{fitting}. For BOSS, we use 1200 DR12 \textsc{Patchy} mocks in each galactic cap. For eBOSS, there are 1000 realizations of \textsc{EZmock} mocks for LRGs in each galactic cap. Both of them are able to reproduce the observational two-point and three-point statistics down to non-linear scales.

\citet{kitaura_modelling_2014} introduces the \textsc{Patchy} code to generate galaxy mocks based on the Augmented Lagrangian Perturbation Theory (ALPT). They determine the galaxy properties by apply their own SHAM to the BigMultiDark simulation\footnote{\url{http://www.multidark.org/}} \citep{bigmultidark} and fit the observational statistics in both real and redshift space. The clustering properties of \textsc{Patchy} mocks are then calibrated by those of SHAM catalogues. The fiducial cosmology is a $\Lambda$CDM cosmology with $\Omega_m=0.307115$, $\Omega_b=0.048206$, $h=0.6777$, $\sigma_8=0.8225$, $n_s=0.9611$ \citep{kitaura_clustering_2016}.

The Effective Zel’dovich approximation mock (\textsc{EZmock}) proposed by \citet{Chuang2015} is another fast methodology to generate mock halo or galaxy catalogues, with the same cosmology as \textsc{Patchy} mocks. \textsc{EZmock} relies on the Zel'dovich approximation \citep{Zeldovich1970} to calculate the displacement of DM particles at any redshift given their initial positions. Tracers are assigned in the density field using both the deterministic and the stochastic bias model \citep{Zhao2020}. Besides covariance matrix calculations, we also use \textsc{EZmock} mocks with and without systematics to study the impact of the biased 2PCF monopole on the projected 2PCF in Section~\ref{wp consistency}. 

\section{Methods}
\label{methods}
\subsection{Galaxy Clustering}
\label{Galaxy Clustering}
The 2PCF, denoted by $\xi$, measures the excess probability of finding a galaxy pair compared to a random distribution in a given volume. We use the Landy--Szalay estimator \citep[LS;][]{Landy1993} which minimises the variances of the measurements: 
\begin{equation}
\label{LS estimator}
   \xi_{\rm LS}  = \frac{\rm DD-2DR+RR}{\rm RR}, 
\end{equation}
where galaxy--galaxy pairs (DD), random--galaxy pairs (DR) and random--random pairs (RR) are normalized by the total number of pairs. $\xi$ and the pair counts can be calculated as a function of the pair separation $s$ and the cosine of the angle between $\textit{\textbf{s}}$ and the line-of-sight ($\beta$), i.e., $\mu=\rm \cos(\beta)$. By weighting the 2D $\xi(s,\mu)$ with Legendre polynomials $P_\ell(\mu)$, we obtain the 1D $\xi$ multipoles as 
\begin{equation}
    \label{multipoles}
    \xi_\ell (s) = \frac{2\ell+1}{2}\int_{-1}^1 \xi(s,\mu) P_\ell(\mu) {\rm d}\mu.
\end{equation}
In our study we only consider the monopole $\xi_0$ and quadrupole $\xi_2$ at $s\in$ [5, 25]$\hmpc$. \mycmt{This is because scales larger than 25$\hmpc$ are prone to imaging systematics \citep{huterer_calibration_2013} while scales smaller than 5$\hmpc$ are sensitive to fibre collision effects. Even with nearest-neighbour close-pair weights $w_{\rm CP}$, the biases of the 2PCF quadrupole due to fibre collision effects can be as large as 3$\sigma$ at $\sim$ 5$\hmpc$ \citep{Guo2012,Rodriguez-Torres2016}. To understand its influence on our SHAM results, we test a fitting range of [10, 25]$\hmpc$ -- which is only mildly affected by fibre collisions -- for the BOSS data. It turns out that the SHAM constraints in this range are consistent with those on $s\in$ [5, 25]$\hmpc$ (see Appendix~\ref{10-25 no bias}). So we use the fitting range $s\in$ [5, 25]$\hmpc$ for all BOSS/eBOSS samples hereafter.}



We use linear $s$ bins with 1$\hmpc$ interval, and 120 linear $\mu$ bins in the range of [0,1), for most of the 2PCF measurements in this work. Only for the eBOSS LRGs in different redshift bins, we have 8 logarithmic bins for $\xi_0$ and $\xi_2$ respectively. 

$\xi$ can also be measured as a function of the distance parallel ($\pi$) and perpendicular ($r_p$) to the line-of-sight. By integrating $\xi(r_p,\pi)$ with respect to $\pi$, one obtains the projected 2PCF as
\begin{equation}
    w_p(r_p) = \int_{-\pi_{\rm max}}^{\pi_{\rm max}}\xi(r_p,\pi){\rm d}\pi.
\end{equation}
We use 8 logarithmic bins for $r_p \in$ [5, 25]$\hmpc$, but $\pi_{\rm max}$ will be changed according to our needs as introduced in Section~\ref{wp consistency}. \textsc{FCFC}\footnote{\url{https://github.com/cheng-zhao/FCFC}} \textcolor{blue}{(Zhao et al. in preparation)} and \textsc{Corrfunc} Python module \citep{10.1007/978-981-13-7729-7_1,2020MNRAS.491.3022S} are employed to calculate $\xi_\ell(s)$ and $w_p$. $\xi_\ell(s)$ of eBOSS are computed with PIP+ANG-weighted pair counts \citep{Mohammad_2020}.

Our SHAM galaxy catalogue is built from an $N$-body simulation in a box, so we convert the position of SHAM galaxies from real space to redshift space before calculating the 2PCF with \citep{Kaiser1987}
\begin{equation}
    Z_{\rm redshift}=Z_{\rm real}+\frac{v_{\rm pec,Z}}{a(t)H(z)}, 
    \label{real-redshift}
\end{equation} 
where $Z$ is the coordinate and $z$ is the redshift, $v_{\rm pec,Z}$ is the peculiar velocity along the $Z$-axis. The estimator used to calculate the SHAM 2PCF is the Peebles--Hauser estimator \citep{Peebles1974}
\begin{equation}
    \label{PH estimator}
    \xi_{\rm PH} = \frac{\rm DD}{\rm RR}-1,
\end{equation}
where galaxy--galaxy pairs (DD) is also normalized by the total number of pairs. The random--random (RR) counts follow the analytical expression:
\begin{equation}
\label{RR(s,mu)}
    \rm RR = \frac{4\uppi}{3}\frac{\textit{s}_\text{max}^3-\textit{s}_\text{min}^3}{\textit{V}_\text{box}} \frac{1}{\textit{N}_{\mu}},
\end{equation}
where $s_\text{max}$ and $s_\text{min}$ are the boundaries of the separation bins, $V_\text{box}=1\,h^{-3}\,{\rm Gpc}^{3}$ is the volume of the simulation box and $N_{\mu}=120$ is the number of $\mu$ bins. We do not apply weights to SHAM galaxies as there are no observational systematics and radial selection. The $\xi_\ell(s)$ of SHAM galaxies is calculated by the built-in 2PCF calculator in our SHAM implementation code.

\subsection{Redshift Uncertainty}
\label{deltav calculation}
The redshift uncertainty is an inevitable error while measuring the redshift. It mainly affects the quadrupole of our clustering measurements \citep{smith_completed_2020}, since it can be regarded as random peculiar motions along the line-of-sight. As suggested in \citet{bolton_spectral_2012}, the redshift determination pipeline is the primary source of the redshift uncertainty of SDSS LRGs, but it is not the dominant source. Hence, using the error given by the pipeline will underestimate the redshift uncertainty. As a result, we estimate it statistically via repeat observations. 

\begin{figure*}
    \centering
    \includegraphics[width=\linewidth,scale=0.8]{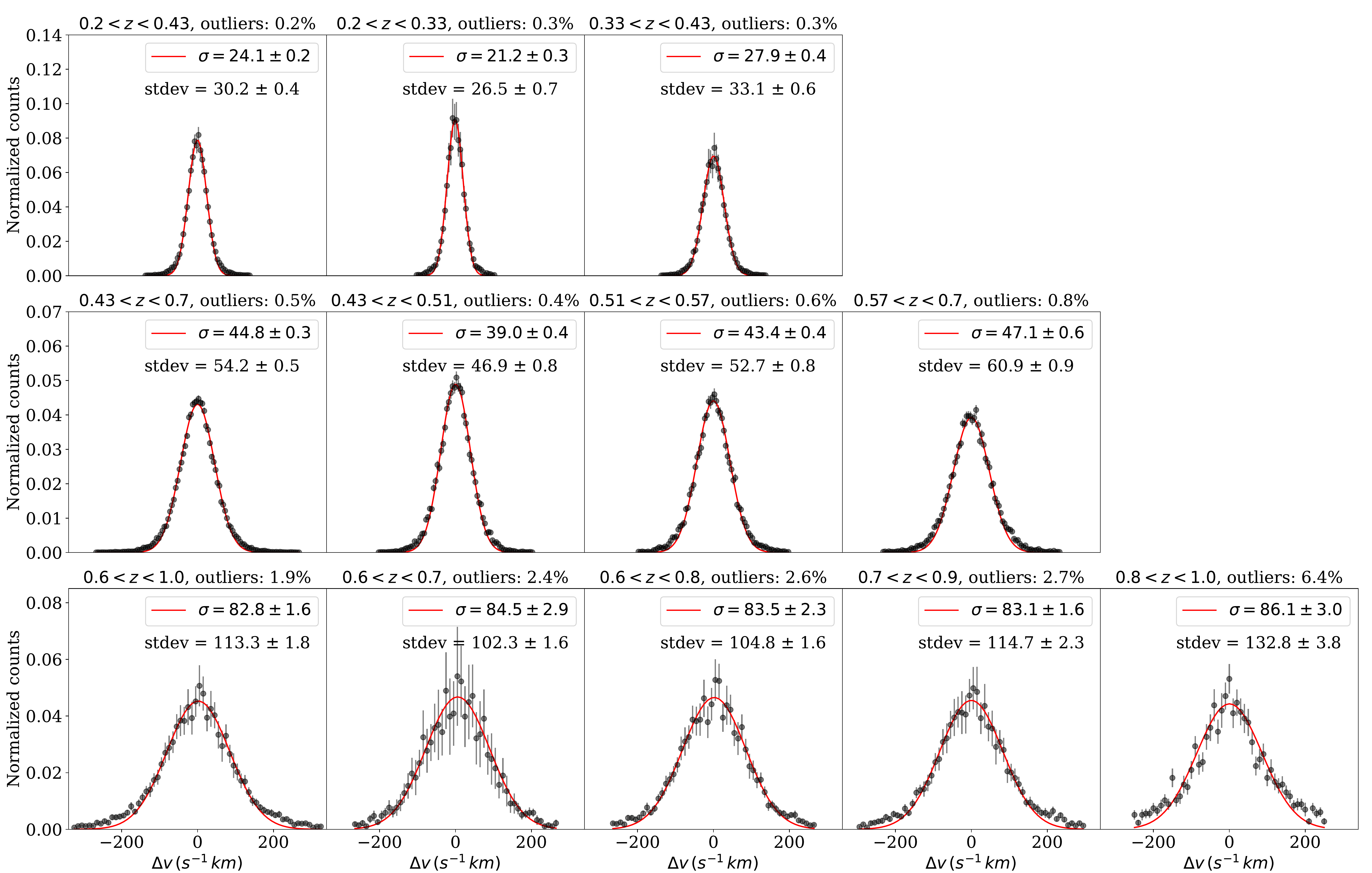}
    \caption{The Gaussian fitting of the $\Delta v$ histograms. The error bars of $\Delta v$ histograms are calculated from jackknife re-sampling. The best-fitting Gaussian models are shown in red lines, labelled with their dispersion values $\sigma_{\Delta v}$. The standard deviations of $\Delta v$ is written below the legend and the proportion of outliers that are beyond the Gaussian fitting range is indicated above the subplots. SDSS has a Gaussian $\Delta v$ histogram, and the ourlier rate of eBOSS are much larger than those of BOSS, consistent with the difference between $\sigma_{\Delta v}$ and $\hat{\sigma}_{\Delta v}$.}
    \label{deltav measurements}
\end{figure*}

The redshift difference between two measurements $\Delta z$ can be converted to $\Delta v$ that represents the radial motion as
\begin{equation}
    \Delta v=\frac{\Delta z}{(1+z)c},
\end{equation}
where $z$ is the redshift of one of measurements with $\textsc{SPECPRIMARY} = 1$. With jackknife errors, we fit all the $\Delta v$ histograms with Gaussian distributions, obtaining the best-fitting dispersion $\sigma_{\Delta v}$ as illustrated in \reffig{deltav measurements}. We also quote their standard deviation, $\hat{\sigma}_{\Delta v}$, as a statistical estimate of the redshift uncertainty. In our SHAM algorithm, its clustering effect is quantified by $v_{\rm smear}$.

\label{Vsmear}

\subsection{SHAM Implementation}
\label{sham implementation}
Recent SHAM studies \citep[e.g.,][]{Cambell2018,Granett2019MNRAS.489..653G,SHAM_hydro_2021} consistently choose $V_{\rm peak}$ to represent the halo mass of haloes and subhaloes. The subsequent the progress of the SHAM algorithm is to sort the galaxy stellar mass and $V_{\rm peak}^{\rm scat}$, which is obtained as
\begin{equation}
\label{SHAM scat}
  V_{\rm peak}^{\rm scat} = V_{\rm peak}S_g,
\end{equation}
where
\begin{equation}
    S_g = 1+\mathcal{N}(0,\sigma^2),
\end{equation}
and match them one by one from the largest $V_{\rm peak}^{\rm scat}$, until the number density of SHAM galaxies equals to that of the observation. In \refeq{SHAM scat}, $\mathcal{N}(0,\sigma^2)$ represents a Gaussian random distribution centred on zero with $\sigma^2$ variance, where $\sigma$ is the only free parameter in the SHAM. \mycmt{The choice of the Gaussian function is due to the fact that the Tully--Fisher relation, i.e., the galaxy--halo mass relation, has Gaussian distributed residuals \citep{willick_homogeneous_1997,steinmetz_cosmological_1999}. We assert 
\begin{equation}
    S_g = \exp(\mathcal{N}(0,\sigma^2)).
\end{equation}
when the Gaussian scatter is negative. Nevertheless, it does not change the fitting results. }

There are four prerequisites for the standard SHAM described above: 1) the cosmology of the simulation has to be close to the true one; 2) the simulation should have a good resolution to resolve the subhaloes down to the low-mass end and halo properties should be accurate; 3) the observed clustering measurements should be unbiased; 4) the observational stellar mass function (SMF) has to be complete in the massive end. If one of these prerequisites is missing, SHAM will not model the observed clustering signal accurately on all scales. 

For eBOSS LRGs, there is a rough truncation of the stellar mass in the massive end (Section~\ref{target selection}). A straightforward way to account for this massive-end incompleteness is to discard the most massive haloes. We introduce a new parameter $V_{\rm ceil}$ to remove the $V_{\rm ceil}$ per cent of the largest scattered $V_{\rm peak}$ (i.e., $V^{\rm scat}_{\rm peak}$). Since $\sigma$ and $V_{\rm ceil}$ both modulate the amplitude of 2PCF monopole, they are degenerate. 

The redshift uncertainty can add bias to the observed 2PCF quadrupole. Therefore, we introduce an extra Gaussian-random distribution $\mathcal{N}(0,v_{\rm smear}^2)$ in the peculiar velocity of SHAM galaxies as
\begin{equation}
\label{vpec smear}
v^{\rm scat}_{\rm pec,Z} =v_{\rm pec,Z} + \mathcal{N}(0,v_{\rm smear}^2).
\end{equation}
The target is to mimic the impact of the measured Gaussian redshift uncertainties in the redshift-space clustering as presented in ~\reffig{deltav measurements}.

To conclude, our 3-parameter SHAM implementation is as follows:
\begin{enumerate}
  \item Scatter the $V_{\rm peak}$ as suggested in \refeq{SHAM scat} with 
  \begin{equation}
        S_g =  \left\{
    \begin{array}{ll}
      1+\mathcal{N}(0,\sigma^2), & \mathcal{N}(0,\sigma^2)\geq 0 \\
      \exp(\mathcal{N}(0,\sigma^2)), &  \mathcal{N}(0,\sigma^2) < 0; \\
    \end{array} 
    \right. 
  \end{equation}
  \item Sort $V^{\rm scat}_{\rm peak}$ and discard the most massive $V_{\rm ceil}$ per cent of haloes/subhaloes;
  \item From the remaining catalogue, keep the $N_{\rm gal}$-th largest $V^{\rm scat}_{\rm peak}$;
  \item Assign galaxies in the centre of those haloes/subhaloes, and smear the peculiar velocities of galaxies along the line of sight with \refeq{vpec smear};
  \item Convert the galaxy coordinates from the real space to the redshift space using \refeq{real-redshift}.
\end{enumerate}

To model correctly the observed clustering, the redshift of the UNIT snapshot should be close to the effective redshift of the data computed with 
\begin{equation}
    z_{\rm eff} = \frac{\sum z w^2}{\sum w^2}.
\end{equation}
where $w$ is the total galaxy weight. For eBOSS \citep{ross_completed_2020},
\begin{equation}
    w = w_{\rm photosys}w_{\rm FKP}w_{\rm noz}w_{\rm CP}, 
\end{equation}
while for BOSS the total weight has a different form \citep{reid_sdss-iii_2016}
\begin{equation}
    w = w_{\rm photosys}w_{\rm FKP}(w_{\rm noz}+w_{\rm CP}-1).
\end{equation}
Then we apply SHAM on this snapshot to find the best-fitting parameters. Besides the 2PCF calculator mentioned in Section~\ref{Galaxy Clustering}, our code also has a built-in \textsc{MultiNest} sampler which will be introduced in Section~\ref{fitting}.

The number of SHAM galaxies $N_{\rm gal}$ is determined by the observed effective number density $n_{\rm eff}$ in a given redshift range calculated as
\begin{equation}
\begin{split}
    n_{\rm eff} &= \sqrt{ \frac{\int n(z)^2 {\rm d}V} { \int {\rm d}V }} 
    = \sqrt{\frac{\int n(\chi)^2 \chi^2 {\rm d}\chi } { \int \chi^2 {\rm d}\chi }},
\end{split}
\end{equation}
where the second equality is due to ${\rm d}V = A_\text{eff}((\chi + {\rm d}\chi)^3 - \chi^3)/3 = A_\text{eff}\chi^2 {\rm d}\chi +\mathcal{O}({\rm d}\chi^2)$, $A_\text{eff}$ is the effective area of the footprint and $\chi$ is the comoving distance for an object at redshift $z$. The values are reported in \reftab{snapshot}. 
\begin{figure}
    \centering
    \includegraphics[scale=0.4]{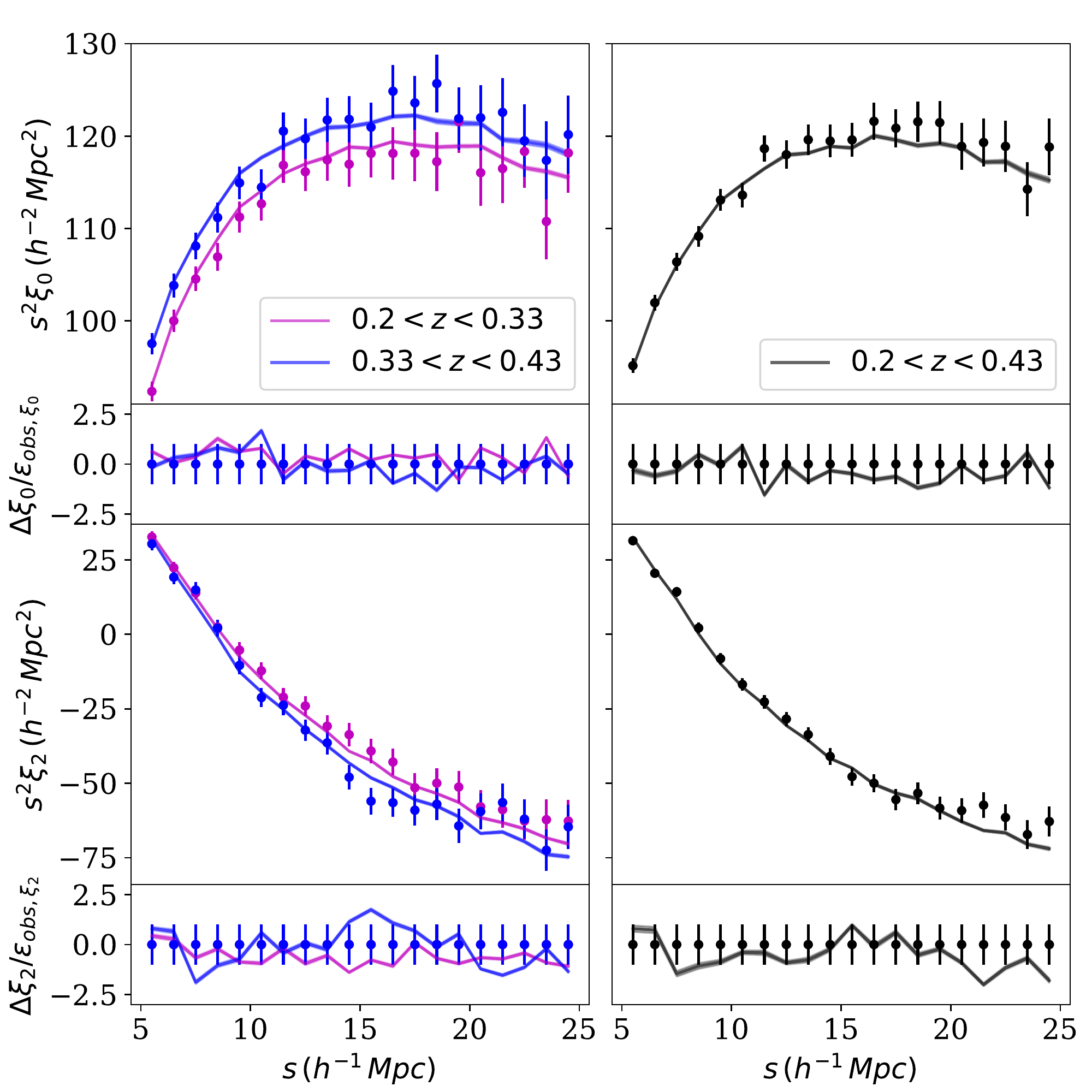}
    \caption{The best-fitting 3-parameter SHAM 2PCF multipoles compared with those of the LOWZ observations on 5--25$\hmpc$. The observational data and the cosmic variance $\epsilon_{\rm obs,\xi_{0,2}}$ are represented by dots with error bars. Solid lines with shades are SHAM 2PCFs and their errors. The first and third row are the monopole and the quadrupole respectively. Residuals normalized by $\epsilon_{\rm obs,\xi_{0,2}}$ are shown in the second and fourth row. The left column is for galaxy samples in redshift slices and the right column is for those in the bulk redshift range. SHAM 2PCFs agree with the observations with rescaled reduced $\chi^2$ values around 1.}
    \label{best-fit lowz}
\end{figure}
\begin{figure}
    \centering
    \includegraphics[scale=0.4]{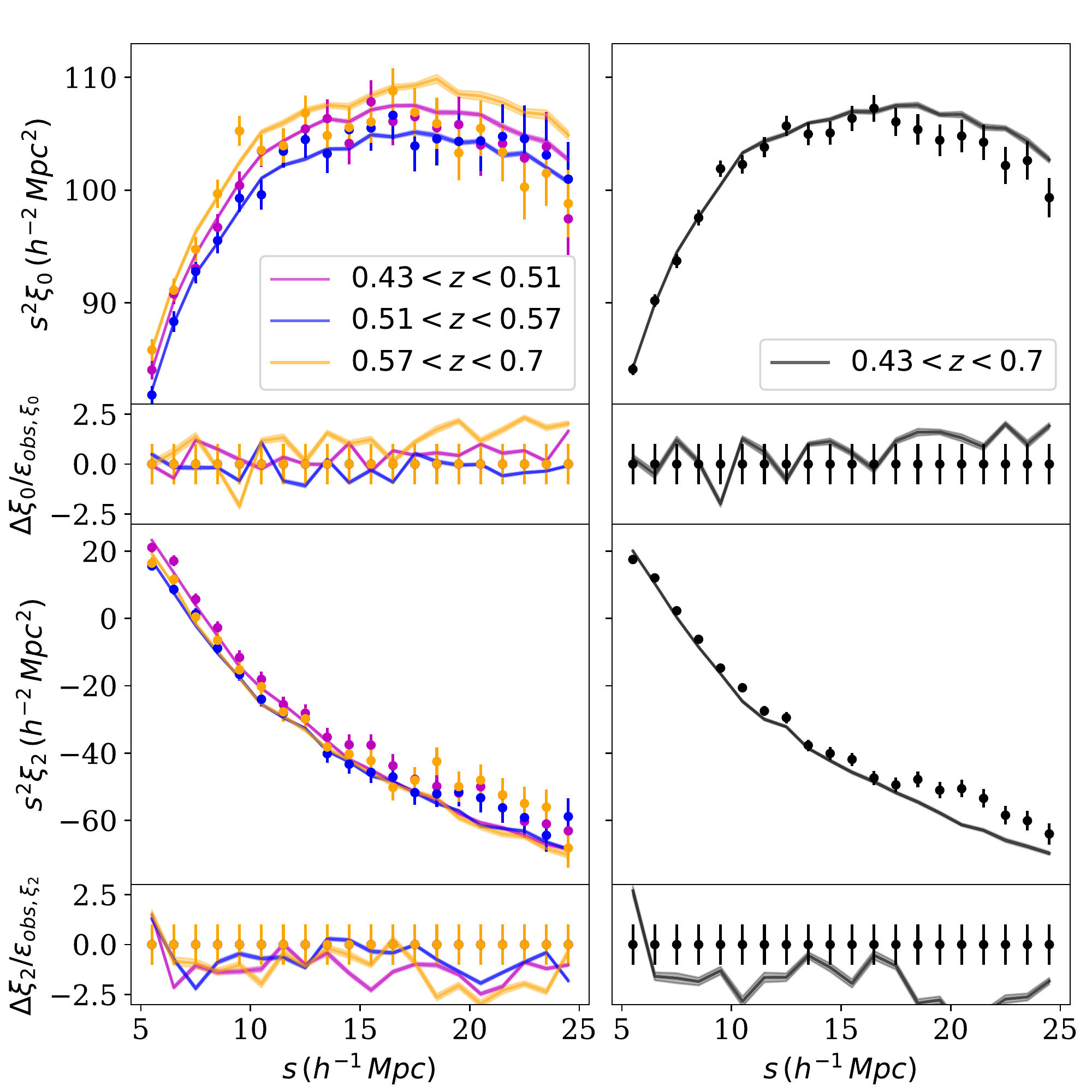}
    \caption{Same figure as \reffig{best-fit lowz}, but for CMASS LRGs. The deviation for the bulk CMASS sample (right panels) is relatively large compared to those of the sliced redshift bins. We attribute this difference to its inhomogeneous sample completeness shown in Section~\ref{result fitting}.}
    \label{best-fit cmass}
\end{figure}
\begin{figure}
    \centering
    \includegraphics[scale=0.4]{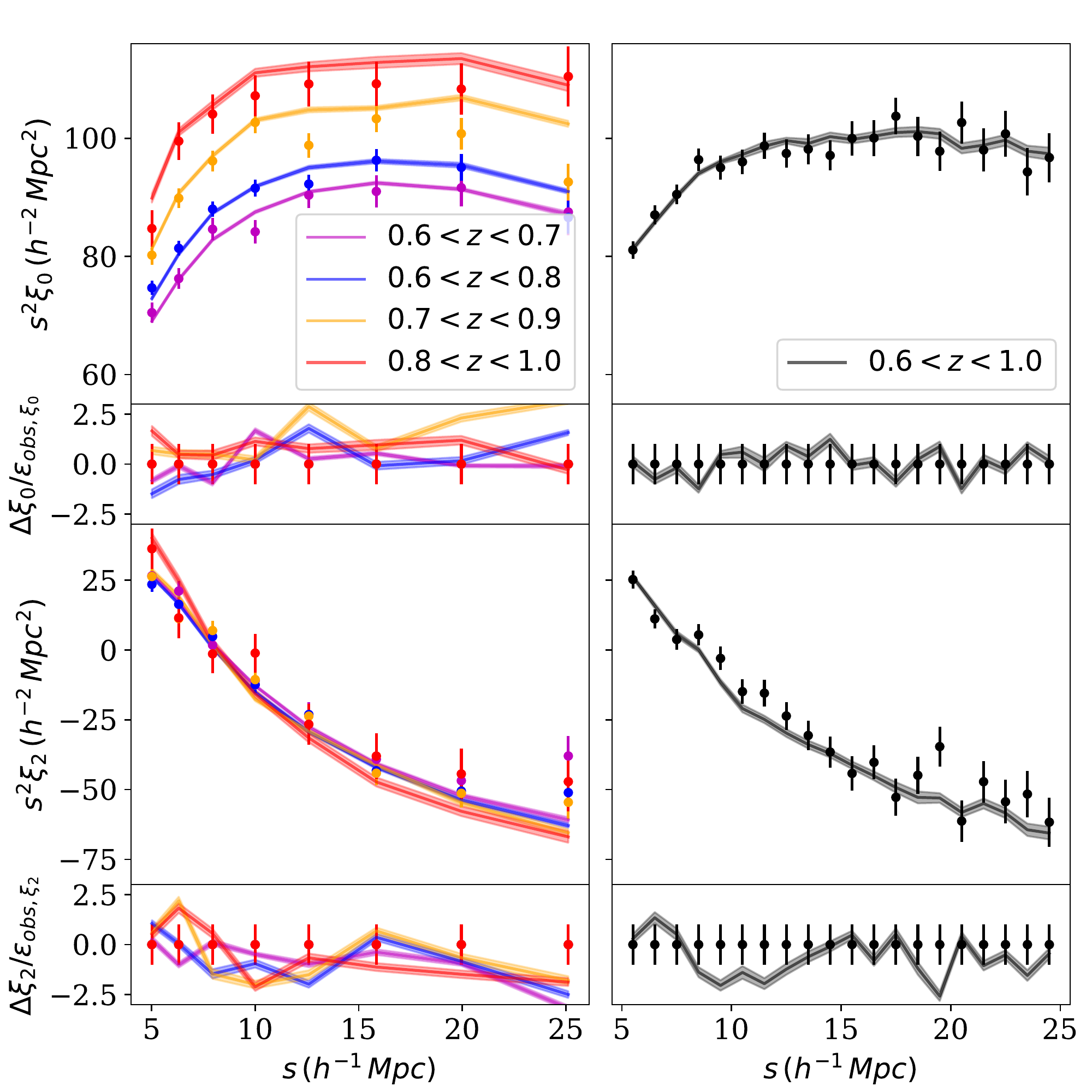}
    \caption{Same figure as \reffig{best-fit lowz}, but for eBOSS LRGs. The SHAM 2PCFs also agree with the observations except the one at $0.7<z<0.9$.}
    \label{best-fit eboss}
\end{figure}

\begin{table*}
{
\begin{tabular}[c]{|c|c|c|c|c|c|c|c|c|c|c|}
{project}&{redshift}    & {$z_{\rm eff}$} & {$z_{\rm UNIT}$} &{ $V_{\rm eff}$}& {10$^4n_{\rm eff}$} & {$\sigma$} &  {$v_{\rm smear}$} & {$V_{\rm ceil}$} &{$\chi^2$/dof} & {rescaled }\\
{}&{range}    & {} & {} &{ $(h^{-3}\,{\rm Gpc}^{3})$}& {$(h^3\,{\rm Mpc}^{-3})$} & {}  & {$({\rm s}^{-1}\,{\rm km})$} &{$(\%)$}& {} & {$\chi^2$/dof}\\
\hline
\hline
\multirow{2}{3em}{LOWZ}&\multirow{2}{7em}{$0.2< z< 0.33$} & \multirow{2}{3em}{{0.2754}} & \multirow{2}{3em}{{0.2760}} & \multirow{2}{2em}{{0.29}}& \multirow{2}{2em}{{3.37}} &{$0.09^{+0.06}_{-0.06}$}&{$100.3^{+7.6}_{-10.4}$}&{$0.0067^{+0.0019}_{-0.0025}$}  &{32/37}&{31/37}\\
&&&&&&{$0.27^{+0.02}_{-0.02}$}&{$66.1^{+8.2}_{-9.8}$}&{/}  &{33/38} &{32/38}\\
\hline
\multirow{2}{3em}{LOWZ}&\multirow{2}{7em}{$0.33<z<0.43$} & \multirow{2}{3em}{0.3865} & \multirow{2}{3em}{0.3941} &\multirow{2}{2em}{0.33} & \multirow{2}{2em}{2.58}&{$0.20^{+0.05}_{-0.05}$}&{$73.9^{+12.2}_{-12.3}$}&{$0.0059^{+0.0026}_{-0.0028}$}  & {51/37}&{50/37} \\
&&&&&&{$0.34^{+0.02}_{-0.01}$}&{$30.5^{+16.7}_{-16.2}$} &{/} & {54/38} & {52/38}\\
\hline
\multirow{2}{3em}{CMASS}&\multirow{2}{7em}{$0.43<z<0.51$}   & \multirow{2}{3em}{0.4686} & \multirow{2}{3em}{0.4573} &\multirow{2}{2em}{0.47} & \multirow{2}{2em}{3.42}   &{$0.30^{+0.05}_{-0.08}$}&{$45.0^{+20.6}_{-16.6}$}&{$0.0078^{+0.0074}_{-0.0040}$} & {41/37}&{39/37}\\
&&&&&&{$0.43^{+0.01}_{-0.02}$}&{$17.4^{+11.8}_{-9.6}$} & {/}& {45/38}& {43/38}\\
\hline
\multirow{2}{3em}{CMASS}&\multirow{2}{7em}{$0.51<z<0.57$}    & \multirow{2}{3em}{0.5417}  & \multirow{2}{3em}{0.5574} &\multirow{2}{2em}{0.46}   & \multirow{2}{2em}{3.63}   &{$0.23^{+0.04}_{-0.04}$}&{$23.2^{+18.5}_{-15.0}$}&{$0.0144^{+0.0037}_{-0.0036}$} & {43/37}   &{41/37} \\
&&&&&&{$0.42^{+0.02}_{-0.01}$}&{$6.4^{+6.2}_{-4.0}$} & {/} & {60/38} & {58/38}   \\
\hline
{CMASS}&{$0.57<z<0.7$}     & {0.6399}  & {0.6281} &{0.65} & {1.60}   &{$0.17^{+0.13}_{-0.02}$}&{$73.1^{+8.2}_{-25.3}$}&{$0.0459^{+0.0029}_{-0.0143}$}  & {54/37} &{51/37}     \\
\hline
{eBOSS}&{$0.6<z<0.7$}     &{0.6518}   & {0.6644}  &{0.16} & {0.939}  &{$0.58^{+0.29}_{-0.22}$}&{$100.4^{+12.3}_{-12.8}$}&{$0.0510^{+0.0324}_{-0.0194}$} & {16/13}   &{16/13}    \\
\hline 
{eBOSS}&{$0.6<z<0.8$}    & {0.7071}      & {0.7018}  &{0.33}     & {0.886}  &{$0.38^{+0.31}_{-0.13}$}&{$98.8^{+10.8}_{-13.4}$}&{$0.0617^{+0.0185}_{-0.0308}$}  & {24/13}  & {23/13}  \\
\hline
{eBOSS}&{$0.7<z<0.9$}   & {0.7968}  & {0.8188} &{0.26}      & {0.647}    &{$0.09^{+0.17}_{-0.04}$}&{$128.5^{+9.5}_{-22.2}$}&{$0.0690^{+0.0044}_{-0.0172}$}   & {30/13}   &{30/13}  \\
\hline
{eBOSS}&{$0.8<z<1.0$}   & {0.8778}& {0.9011}    &{0.09} & {0.301}    &{$0.22^{+0.17}_{-0.14}$}&{$134.2^{+18.5}_{-20.1}$}&{$0.0481^{+0.0153}_{-0.0155}$}  & {16/13}  &{16/13}   \\
\hline
\hline
\multirow{2}{3em}{LOWZ}&\multirow{2}{7em}{$0.2<z<0.43$}    & \multirow{2}{3em}{0.3441}& \multirow{2}{3em}{0.3337}  &\multirow{2}{2em}{0.62}    & \multirow{2}{2em}{2.95}     &{$0.22^{+0.03}_{-0.05}$}&{$76.6^{+10.2}_{-8.2}$}&{$0.0031^{+0.0023}_{-0.0015}$} & {45/37}  &{42/37}   \\
&&&&&&{$0.30^{+0.01}_{-0.01}$}&{$54.5^{+6.9}_{-6.7}$} & {/}&{45/38} & {42/38}\\
\hline
{CMASS}&{$0.43<z<0.7$}  & {0.5897} & {0.5924} &{1.58}     & {2.64}    &{$0.20^{+0.20}_{-0.03}$}&{$63.5^{+9.1}_{-44.6}$}&{$0.0269^{+0.0032}_{-0.0190}$} & {81/37}   &{70/37}    \\
\hline
{eBOSS}&{$0.6<z<1.0$}  & {0.7781}  & {0.7018}  &{0.43}    & {0.626}    &{$0.40^{+0.19}_{-0.06}$}&{$109.7^{+8.0}_{-6.1}$}&{$0.0542^{+0.0067}_{-0.0196}$}   & {35/37}   &{33/37}    \\
\hline
\end{tabular}}
\caption{\mycmt{Properties of analysed galaxy samples and their best-fitting SHAM results. The first six columns are the project names of galaxy samples, their redshift ranges, effective redshifts $z_{\rm eff}$, the closest redshifts of the UNIT snapshot to $z_{\rm eff}$, effective volumes, and their average number densities. The next four columns are the constraints of SHAM parameters \{$\sigma, v_{\rm smear}, V_{\rm ceil}$\} and the minimum $\chi^2$/dof. The final column is the rescaled $\chi^2$/dof taking into account the uncertainty of UNIT simulations. There are redshift bins with two sets of SHAM parameters. The first set (row) is from the 3-parameter SHAM and the second set is from 2-parameter SHAM without $V_{\rm ceil}$. The 2-parameter SHAM can only be applied to complete samples, i.e., LOWZ and CMASS galaxies at $z<0.6$ \citep{leauthaud_stripe_2016}. Redshift bins with only one set of SHAM results are from the 3-parameter SHAM. }
}
\label{snapshot}
\end{table*}

\subsection{SHAM Fitting}
\label{fitting}
The SHAM best-fitting parameters are obtained by minimizing the $\chi^2$ value (i.e., the maximum log-likelihood) between the SHAM and the observational 2PCF. The $\chi^2$ for a given parameter set $\Theta=\{\sigma, v_{\rm smear},V_{\rm ceil}\}$ is defined as 
\begin{equation}
\chi^2 (\Theta) =  (\xi_{\rm data}-\xi_{\rm model}(\Theta))^T\vb*{C}^{-1}(\xi_{\rm data}-\xi_{\rm model}(\Theta)),
\end{equation}
where $\xi = (\xi_0,\xi_2)$ denotes the vector composed of the 2PCF monopole and quadrupole. The vectors $\xi_{\rm data}$ and $\xi_{\rm model}$ represent the data vector and the SHAM model 2PCF respectively. In particular, $\xi_{\rm model}$ is obtained by averaging the 2PCFs of 32 SHAM galaxy realizations generated using the same $\Theta$ with different random seeds. This operation reduces the SHAM statistical uncertainty. $\vb*{C}$ is the unbiased covariance matrix \citep{Hartlap2007}: 
\begin{equation}
    \vb*{C}^{-1} = \vb*{C}^{-1}_{s} \frac{N_{\rm m}-N_{s}-2}{N_{\rm m}-1},
\end{equation}
where $N_{s}$ is the length of $\xi_{\rm data}$, i.e., the total number of bins used in the 2PCF fitting, and $N_{\rm m}$ is the number of mocks used to compute the covariance matrix. $N_{\rm m}=1200$ for BOSS \textsc{Patchy} mocks and $N_{\rm m}=1000$ for eBOSS \textsc{EZmock} mocks. $\vb*{C}_{s}$ is the covariance matrix calculated as \citep{Zhao2020} 
\begin{equation}
   \vb*{C}_{s,\rm ij} = \frac{1}{\rm N_{\rm m}-1}\sum^{\rm N_{\rm m}}_{k=1} [\xi_{\rm k}(s_{\rm i})-\overline{\xi}(s_{\rm i})][\xi_{\rm k}(s_{\rm j})-\overline{\xi}(s_{\rm j})],
\end{equation}
where $\xi_{\rm k}$ is the correlation function measured from the $k_{\rm th}$ mock, and $\overline{\xi}(s_{\rm i})$ is the average of the mock correlation function in a given distance bin $ s_{\rm i}$. 

We assume a Gaussian likelihood $\mathcal{L}(\Theta)$ for our parameters, which is 
\begin{equation}
\label{likelihood}
    \mathcal{L}(\Theta) \propto \rm e^{-\frac{\chi^2(\Theta)}{2}},
\end{equation}
and employ a Monte-Carlo sampler \textsc{Multinest}\footnote{\url{https://github.com/farhanferoz/MultiNest}} \citep{Feroz2008,2009MNRAS.398.1601F,2019OJAp....2E..10F}, an efficient nested sampling technique especially for multi-modal posteriors, to constrain $\Theta$. The survival volume (i.e., the prior volume) is defined as \citep{Feroz2008}
\begin{equation}
    X(\lambda) = \int_{\{\Theta: \mathcal{L}(\Theta)>\lambda\}}\text{Pr}(\Theta|H){\rm d}\Theta,
\end{equation}
where $\text{Pr}(\Theta|H)$ is the parameter prior, and the evidence integral can be written as 
\begin{equation}
    \mathcal{Z} = \int_0^1 \mathcal{L}(\Theta) {\rm d}X \approx \sum^M_{i=1}\mathcal{L}_iw_i,
\end{equation}
where $i$ is the number of iterations, $w_i = \frac{1}{2}(X_{i-1}-X_{i+1})$, and $X_i$ is a sequence of decreasing values as $0<X_M<...<X_2<X_1<X_0=1$. During the sampling, live points walk randomly and simultaneously in the parameter space confined by the prior, and some of them will be deactivated if they have the lowest likelihood $\mathcal{L}_i$. The sampling terminates if the evidence contribution from the $j_{\rm th}$ iteration $\Delta\mathcal{Z}_j = \mathcal{L}_{\rm max}X_i$ is lower than a certain threshold (i.e., the tolerance), where $\mathcal{L}_{\rm max}$ is the maximum likelihood among the current set of live points \citep{2019OJAp....2E..10F}. The final results of our SHAM do not change much when changing the tolerance and the initial number of live points. Therefore, for all the tests, we set the tolerance to be 0.5 and the number of particles to be 200, in order to improve the efficiency of the convergence. 

The built-in analyser of \textsc{pyMultinest}\footnote{\url{https://github.com/JohannesBuchner/PyMultiNest}} \citep{pymultinest} is used to extract the median value and the 1$\sigma$ limits of the parameters, and the maximum-likelihood $\chi^2$ listed in \reftab{snapshot}. The parameter set with the maximum-likelihood $\chi^2$ is indicated in the posteriors provided by  \textsc{Getdist} \citep{2019arXiv191013970L} in Appendix~\ref{appendix posteriors}. 

\section{Results}
\label{Results}
\begin{figure*}
    \centering
    \includegraphics[scale=0.5]{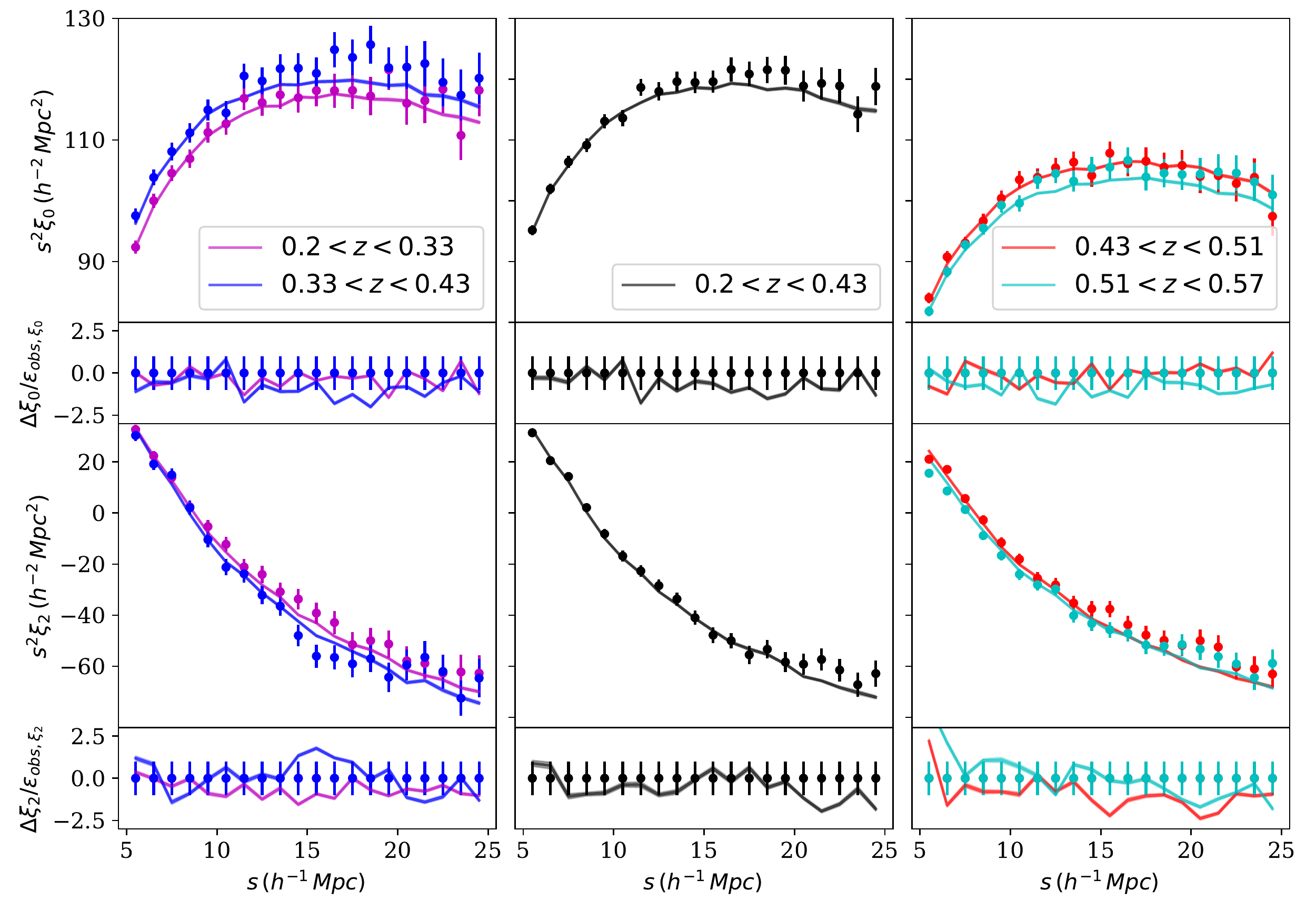}
    \caption{2PCF multipoles of the best-fitting 2-parameter SHAM (with $\sigma$ and $v_{\rm smear}$) compared with those of the observations on 5--25$\hmpc$ as in \reffig{best-fit lowz}. The first column is for LOWZ in redshift slices, the second column is for LOWZ at a bulk redshift range, and the third column is for CMASS in redshift slices. SHAM 2PCF at $z<0.51$ agrees with the observational 2PCF. However, a significant difference in the first two bins of the quadrupole is shown for the CMASS sample at $0.51<z<0.57$.}
    \label{best-fit 2-param}
\end{figure*}
\subsection{Galaxy Clustering Fitting}
\label{result fitting}
We calibrate our SHAM model with LRGs on 5--25$\hmpc$ for the full LOWZ, CMASS and eBOSS LRG samples as well as 9 finer redshift bins as summarized in \reftab{snapshot}. Figures~\ref{best-fit lowz}--\ref{best-fit eboss} show that 2PCFs of the SHAM catalogues agree with those of the observational data in general. Note that there are 32 realizations of SHAM catalogues produced by the maximum-likelihood parameter set, aiming at reducing the statistical error brought by the Gaussian scatters. For this reason, we choose the mean value of their 2PCFs to be the SHAM 2PCF, and rescale their 2PCF standard deviations by $1/\sqrt{32}$ to obtain the error of the SHAM. We find that the SHAM error is negligible compared to observed statistical error determined by \textsc{PATCHY} mocks or \textsc{EZmock} mocks. Thus, we decided to ignore it in this study. 

The reduced $\chi^2$ values from the fits do not include cosmic variances of the UNIT simulations. To include their influence in the analysis, we divide the reduced $\chi^2$ by
\begin{equation}
    \frac{\epsilon^2_{\rm com}}{\epsilon^2_{\rm obs}} = 1+\frac{\epsilon_{\rm UNIT}^2}{\epsilon_{\rm obs}^2} \approx 1+\frac{V_{\rm eff, obs}}{V_{\rm eff,UNIT}},
\end{equation}
where $\epsilon^2_{\rm obs}$ and $\epsilon^2_{\rm UNIT}$ represent the observational variance and the cosmic variance of UNIT simulation respectively and $\epsilon^2_{\rm com}$ is the combination of two errors. $V_{\rm eff, obs}$ is the effective volume calculated with \citep{2013Veff,reid_sdss-iii_2016}
\begin{equation}
    V_{\rm eff, obs} = \sum_i \left(\frac{\bar{n}(z_i)P_0}{1+\bar{n}(z_i)P_0}\right)^2 \Delta V(z_i),
\end{equation}
where $\Delta V(z_i)$ is the volume of the shell at redshift $z_i$ and $\overline{n}(z_i)$ is the mean number density of that shell. The values are listed in the fourth column of \reftab{snapshot} and we assume $V_{\rm eff,UNIT}=10\,h^{-3}\,{\rm Gpc}^{3}$ \citep{2019MNRAS.487...48C}. The rescaled $\chi^2$ values are recorded in the final column of \reftab{snapshot}. As the rescaled value is a better indicator of the goodness of the fit than the $\chi^2$ value, our discussions hereafter are all based on the rescaled ones. 

The $\chi^2$ value for the bulk CMASS sample is relatively large compared to those of CMASS redshift slices and other bulk samples. This can be attributed to the varying sample completeness of CMASS sample which we will prove later in this chapter. The 2PCF quadrupoles of all CMASS SHAM samples are underestimated on 18--25$\hmpc$, which is consistent with the results of \citet{Rodriguez-Torres2016}. Similar (but less obvious) discrepancies are also observed from the eBOSS samples. They may be due to some uncorrected systematics. The monopole disagreement on $r>10\hmpc$ and the resulting large $\chi^2$ value for eBOSS SHAM at $0.7<z<0.9$ might be due to observational systematics as well.

Our measured $\sigma\in [0.17,0.40]$ for CMASS at $0.43<z<0.7$ is consistent with $\sigma=0.31$ obtained by \citet{Rodriguez-Torres2016}. It is worth noting that their scatter considers the incompleteness of the observed stellar mass function, i.e., they obtain the intrinsic scatter $\sigma_{\rm int}$. But we assume that the observed stellar mass function in a certain range is complete. It means that our $\sigma$ account for both the intrinsic scatter and the observed incompleteness.

Since BOSS samples are complete at $z<0.6$ \citep{reid_sdss-iii_2016, leauthaud_stripe_2016}, we also apply our SHAM model without $V_{\rm ceil}$ (2-parameter SHAM hereafter) to the same simulation snapshot as the 3-parameter version and fit to the same data. The best-fitting 2PCFs and parameter constraints are shown in \reffig{best-fit 2-param} and \reftab{snapshot} respectively. The rescaled reduced $\chi^2$ values for the 2- and 3-parameter SHAM models are similar for all LOWZ LRGs and CMASS LRGs at $0.43<z<0.51$, which confirms their galaxy completeness. The reduced $\chi^2$ for the 2-parameter SHAM at $0.51<z<0.57$ is significantly larger (the $\chi^2$ difference shows that SHAM with $V_{\rm ceil}=0$ is rejected more than 4$\sigma$), due to the discrepancy of quadrupole on 5--7$\hmpc$. It also means $\sigma$ and $V_{\rm ceil}$ are not completely degenerate. The monopole of the best-fitting 2-parameter SHAM is also systematically lower than that of the observation. The issue in the result of the 2-parameter SHAM demonstrate that CMASS at $0.51<z<0.57$ is not complete, i.e., $V_{\rm ceil}\neq0$, contrary to what has been presented in \citet{reid_sdss-iii_2016}. This is consistent with the marginalized posterior distributions of $V_{\rm ceil}$ in the 3-parameter SHAM (see \reffig{cmass posterior}).

\subsection{Redshift Uncertainty indicated by \texorpdfstring{$v_{\rm smear}$}{Vsmear}}
\label{vsmear vs repeat}
\begin{figure*}
    \centering
    \includegraphics[scale=0.55]{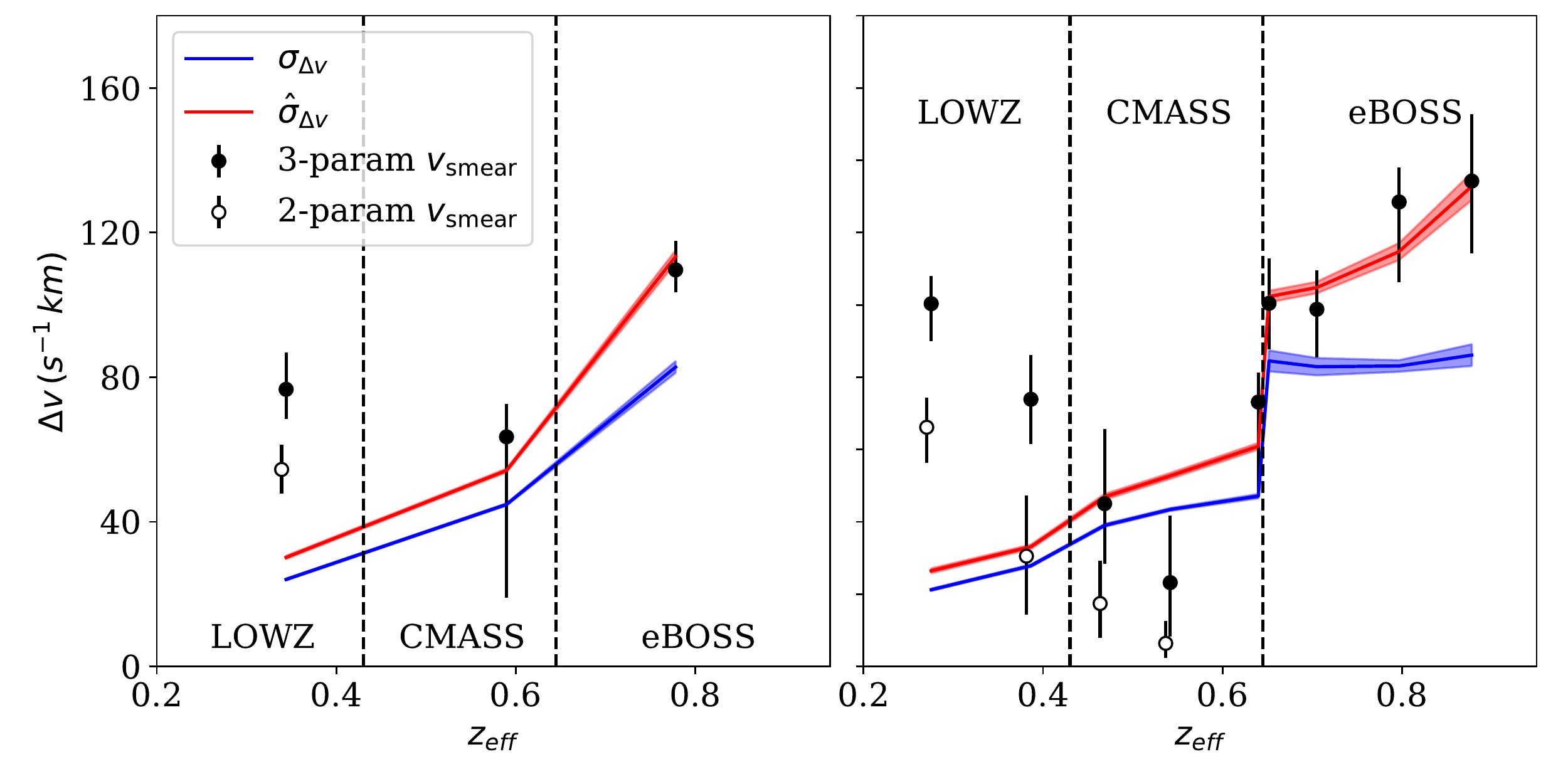}
    \caption{Comparison between the measured $v_{\rm smear}$ from the 3-parameter SHAM (black dots with error bars) and that from the 2-parameter SHAM (open circles with error bars; a horizontal offset is applied to avoid overlapping), the best-fitting Gaussian dispersion $\sigma_{\Delta v}$ (blue line with shades) and the standard deviations $\hat{\sigma}_{\Delta v}$ (red lines with shades) of the $\Delta v$ distribution. The left panel is for the entire sample, while the right panel is for samples in redshift slices. Both the BOSS CMASS sample and the eBOSS LRG sample show good agreements between $v_{\rm smear}$ values and the redshift uncertainties measured from repeat samples. However, for the BOSS LOWZ sample, $v_{\rm smear}$ values are significant larger than those estimated from repeat observations.} 
    \label{vsmear evolve}
\end{figure*}
\begin{figure*}
    \centering
    \includegraphics[scale=0.5]{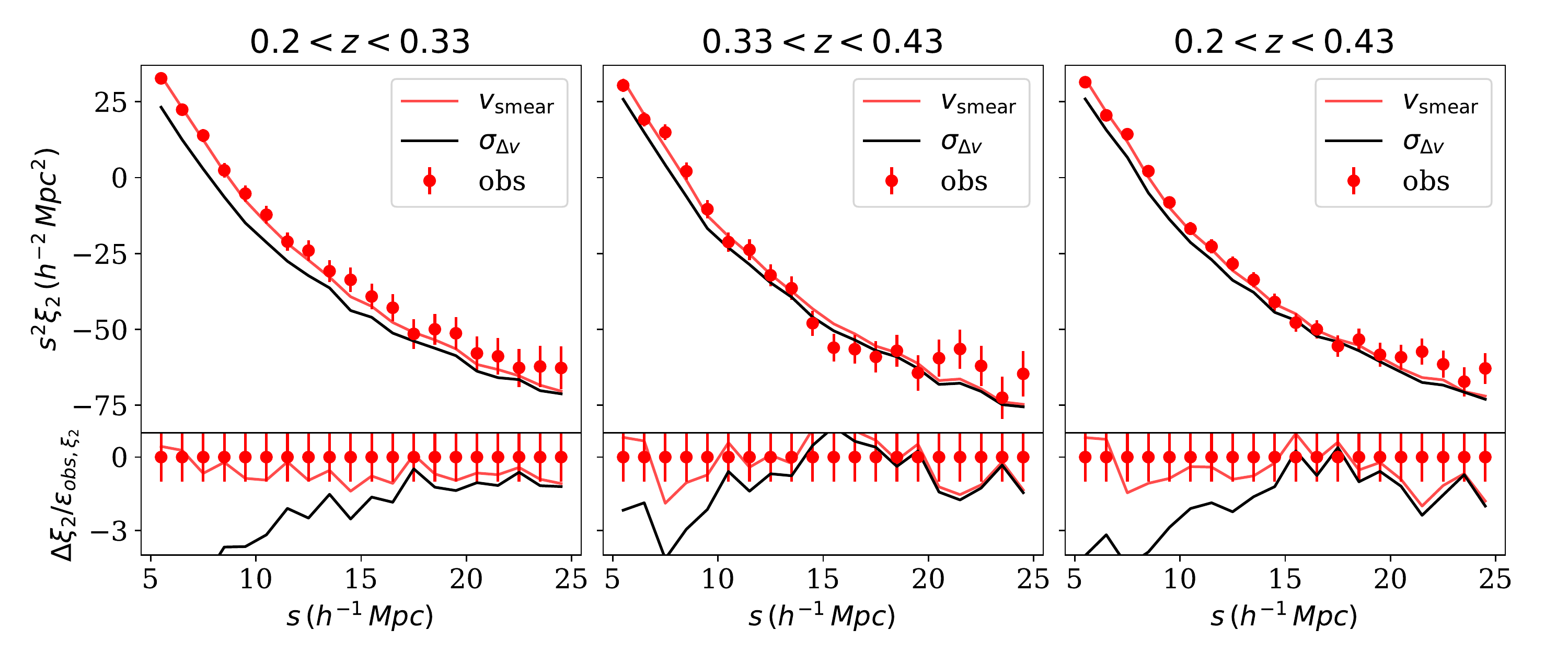}
    \caption{Comparison among the quadrupoles from the observation (red dots with error bars), the best-fitting SHAM (red lines), and the SHAM 2PCF obtained by replacing the best-fitting $v_{\rm smear}$ with $\sigma_{\Delta v}$ (black lines). The first, second and third column represents LOWZ samples at $0.2<z<0.33$, $0.33<z<0.43$ and $0.2<z<0.43$, respectively. The differences between two quadrupoles on $r<10\hmpc$ are mostly larger than 3$\sigma$.}  
    \label{LOWZmps}
\end{figure*}
\begin{figure*}
    \centering
    \includegraphics[scale=0.5]{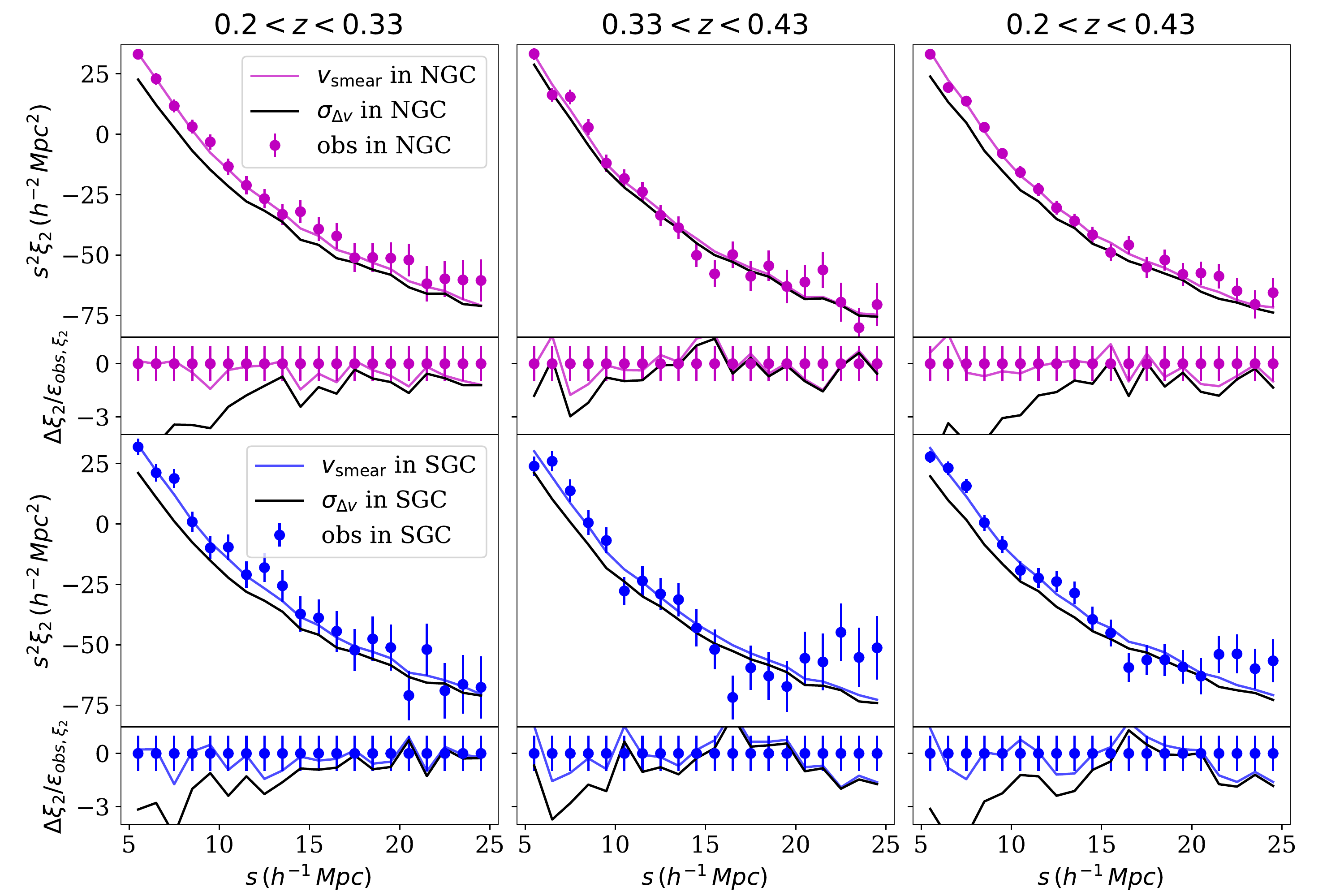}
    \caption{Same as \reffig{LOWZmps}, but using the LOWZ NGC sample (magenta lines) and LOWZ SGC sample (blue lines) separately. The discrepancies between $v_{\rm smear}$ and $\sigma_{\Delta v}$ are not mitigated for LOWZ SGC when there is a high rate of SDSS-III sample.}  
    \label{LOWZNGCSGC}
\end{figure*}

\reffig{vsmear evolve} shows that $\hat{\sigma}_{\Delta v}$ increases monotonically with $z_{\rm eff}$. This is consistent with the fact that the absorption lines used to determine the redshift are broader at higher redshift than those at lower redshift, leading to larger redshift uncertainties. There is a general consistency between the best-fitting SHAM $v_{\rm smear}$ and $\hat{\sigma}_{\Delta v}$, $\sigma_{\Delta v}$ for CMASS. For the eBOSS samples, SHAM $v_{\rm smear}$ agrees with $\hat{\sigma}_{\Delta v}$, but both of them are systematically larger than $\sigma_{\Delta v}$. It means that we cannot neglect the Gaussian-fitting outliers and their effects on eBOSS quadrupoles. The tail of eBOSS $\Delta v$ distributions also slightly deviates from the Gaussian model, meaning that eBOSS $\sigma_{\Delta v}$ is not a good representative of the observed redshift uncertainty. \mycmt{The best-fitting $v_{\rm smear}$ values of 2-parameter SHAM are lower than those of the 3-parameter SHAM, and this difference is larger for complete samples at $z<0.51$ than that of CMASS at $0.51<z<0.57$. This is not a contradiction because the biggest change in $\sigma$--$v_{\rm smear}$ posterior happens in CMASS at $0.51<z<0.57$ as shown in Figures~\ref{2-param LOWZtot}--\ref{2-param CMASS}. The small change in $v_{\rm smear}$ is due to its weaker $V_{\rm ceil}$--$v_{\rm smear}$ degeneracy.
}

Discrepancies also exist for LOWZ LRGs at all redshifts and they remain for the 2-parameter LOWZ SHAM. To compare the differences in the quadrupole, we replace the best-fitting $v_{\rm smear}$ values with $\sigma_{\Delta v}$, generate SHAM galaxy catalogues and calculate the 2PCFs. As presented in \reffig{LOWZmps}, their quadrupole differences are larger than 3$\sigma$. It means that besides the statistical redshift uncertainty, LOWZ has unknown factors that boost the quadrupole. We investigate various potential causes listed below.

\textbf{A. Missing subhaloes or velocity bias in the simulation:}
The host haloes/subhaloes for LRGs have a large mass (e.g. $>1000$ particles), so we expect they are well resolved. But their peculiar velocity might have biases especially for subhaloes. However, our simulation-based SHAM provides $v_{\rm smear}$ that agrees with CMASS and eBOSS samples and we do not expect a significant evolution in the velocity bias. Therefore, the properties of haloes and subhaloes should be reliable and do not lead to the $v_{\rm smear}$ disagreement in LOWZ.

\textbf{B. Observational systematics:}
\label{vsmear discrepancy}
The LOWZ sample includes SDSS-I/II galaxies with a redshift determination pipeline different from that of SDSS-III BOSS \citep{bolton_spectral_2012}. Thus, it is possible that the spectroscopic pipelines yield different uncertainties of redshift measurements. In fact, we find that in the clustering catalogue, 56 per cent of galaxies in the North Galactic Cap (NGC) and 90 per cent of galaxies in the South Galactic Cap (SGC) are from SDSS-III. The component difference in two galactic caps indicates that we can analyse NGC and SGC separately. Repeated samples are all from SDSS-III. Therefore, the $\Delta v$ from repetitive observations may not be representative of the redshift uncertainty of the LOWZ clustering. 

However, as shown in \reffig{LOWZNGCSGC}, a high proportion of SDSS-III galaxies in the SGC does not mean a smaller difference between the $\sigma_{\Delta v}$-generated 2PCF and the $v_{\rm smear}$-generated 2PCF compared to the difference for LOWZ in the NGC. Therefore, the difference in the two spectroscopic pipelines cannot explain the $v_{\rm smear}$ discrepancy.

\textbf{C. Underestimation of redshift uncertainties:}
Considering the fact that errors from the spectroscopic pipeline underestimate redshift uncertainties, it is difficult to directly prove the representative of samples from the repeat observation. Nevertheless, we can proceed with the photometric information (Appendix~\ref{blue-red ratio}). In the LOWZ sample, we find that the fraction of the blue samples $f_{\rm blue}$ in repeat observation is smaller than that of the clustering sample, but the differences are minor (See \reftab{blue ratios cut 1} and \reftab{blue ratios cut 2}). Given the similarity in the $f_{\rm blue}$ and colour--redshift distributions, the sample of the repeat observations are unbiased representatives of galaxies in the clustering catalogue. Even if the repeated samples are biased, this is not the main cause of the discrepancy. Because the redshift uncertainty of LOWZ at lower redshift should be smaller than that of CMASS, but $v_{\rm smear}$ of LOWZ is systematically larger than that of CMASS. 

\textbf{D. A more complete model for galaxy assignment:}
While we are using the same model for CMASS and eBOSS LRGs to describe LOWZ, the discrepancy might indicate that a more complete model is required. E.g., the LOWZ sample is composed of a different type of galaxies which has more satellites than CMASS.

In fact, our satellite fraction ($f_{\rm sat}$) for LOWZ at $0.2<z<0.43$ is 12.6 per cent, similar to CMASS $f_{\rm sat}=12.3$ per cent at $0.43<z<0.7$. They are close to results from Halo Occupation Distribution fitting with $w_p$, another empirical model for the galaxy--halo relation: \citet{2013LOWZsat} based on DR9 give $12\pm 2$ per cent for LOWZ at $0.2<z<0.4$ and \citet{2014CMASSfsat} based on DR10 give $10\pm 2$ per cent at $0.43<z<0.7$. Both SHAM and HOD do not support the explanation of large $f_{\rm sat}$ in LOWZ. Nevertheless, given the difference in data and fitting scales, $f_{\rm sat}$ can still be the explanation. Additionally, as pointed out by \citet{ross_clustering_2014} and \citet{favole_building_2016}, we know that the blue tail of CMASS galaxies show a higher quadrupole at 5--40$\hmpc$ compared to the dominating redder galaxies, which has 2PCF close to that of the full catalogue. For LOWZ, the ad hoc blue galaxies might play a more important role in the clustering properties. The velocity bias is also found to be different in CMASS and LOWZ samples \citep{velocityibas_CMASS, velocitybias_LOWZ}. So it is possible that the LOWZ discrepancy can be mitigated by a better model that considers those factors. We leave it for a future work.

\begin{figure}
    \centering
    \includegraphics[width=\linewidth]{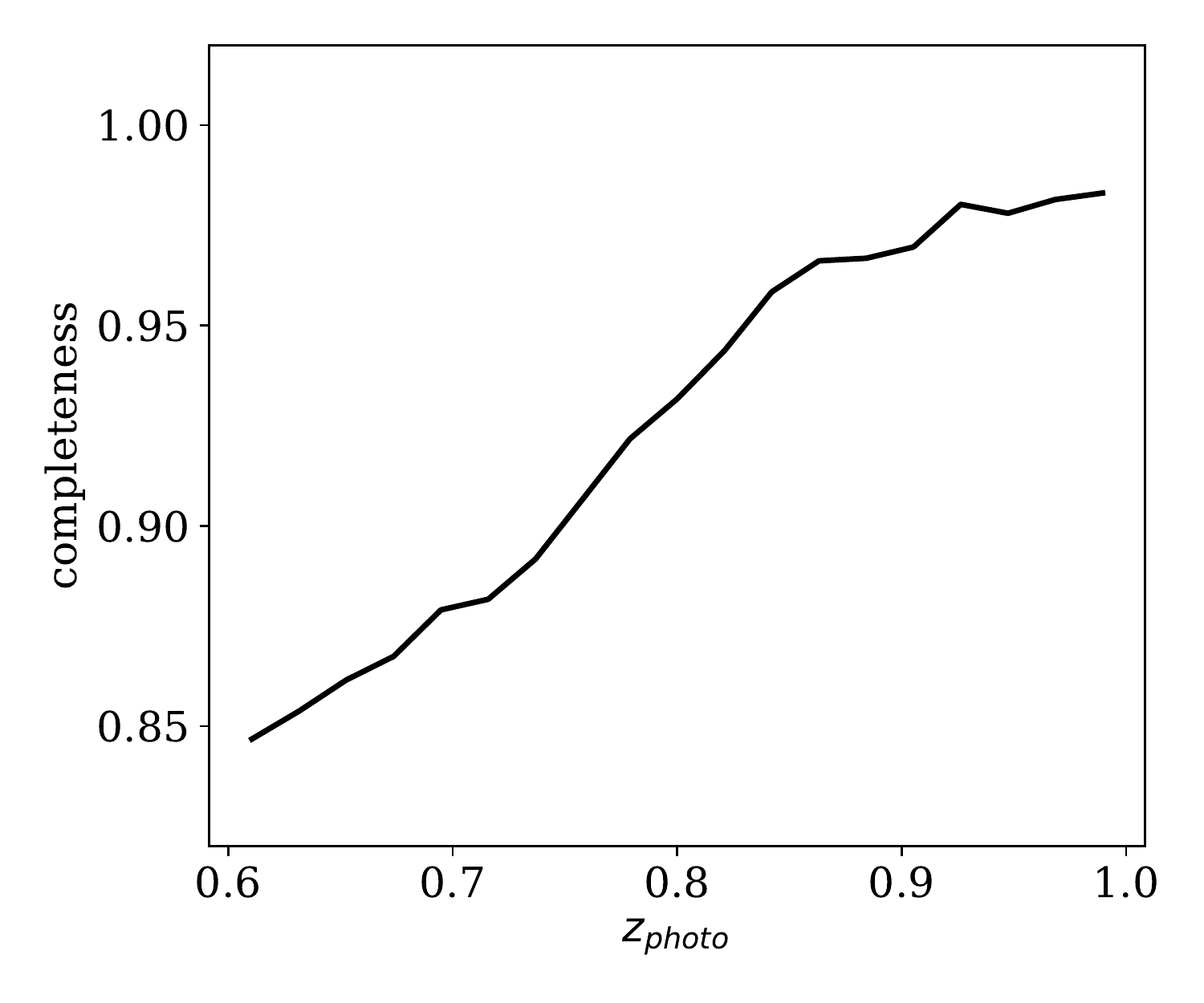}
    \caption{The redshift evolution of the completeness of eBOSS LRG targets. The photometric redshifts ($z_{\rm photo}$) are taken from the DECaLS DR9 catalogues. The increasing completeness with respect to the redshift is the result of the i-band lower limit imposed in the LRG target selection. }
    \label{comp eboss}
\end{figure}
\begin{figure}
    \centering
    \includegraphics[width=\linewidth]{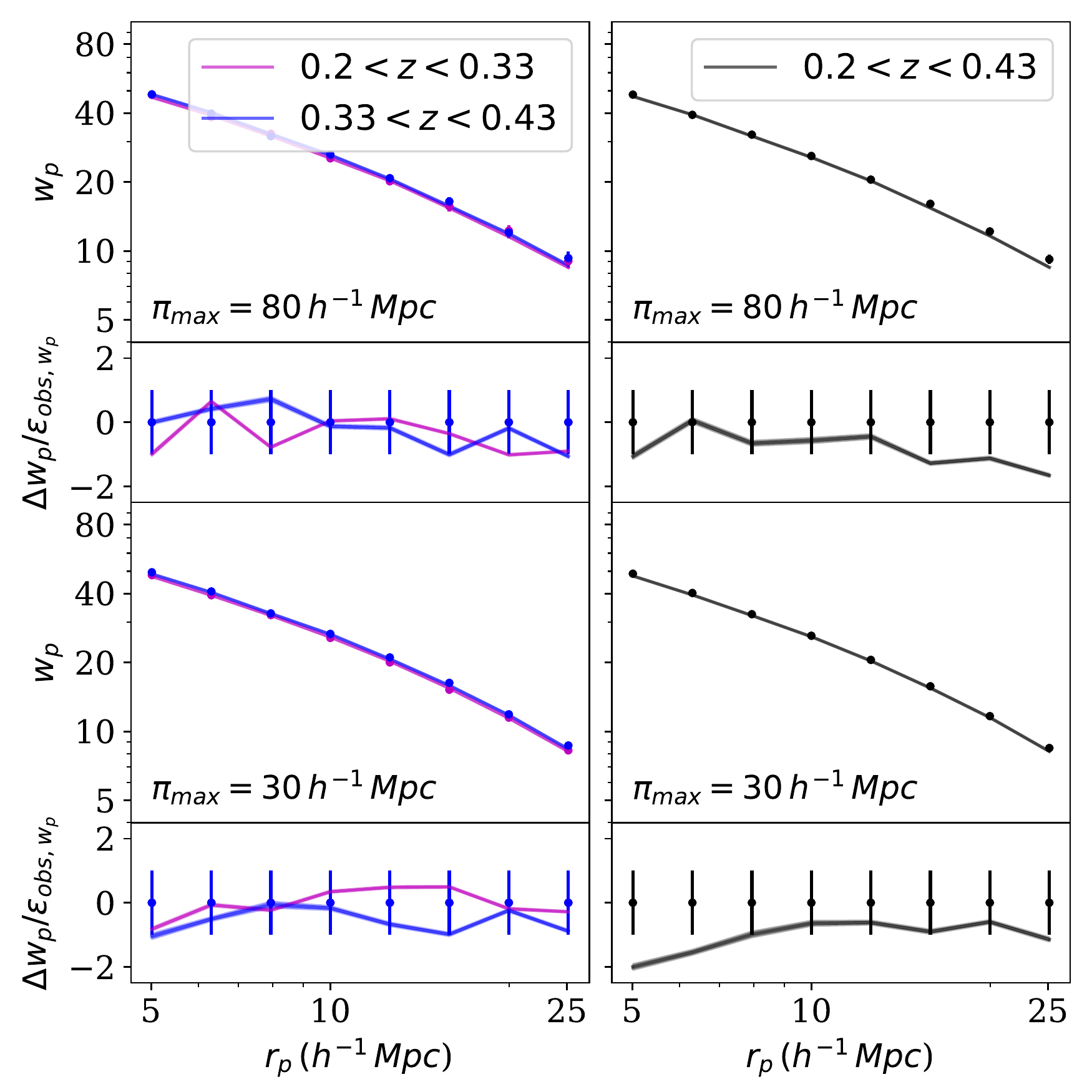}
    \caption{The $w_p$ comparison between the best-fitting SHAM catalogues and the observations. The first row and the third row are $w_p$ with $\pi$ integrating up to 80$\hmpc$ and 30$\hmpc$ respectively, and their residuals normalized by the cosmic variance $\epsilon_{\rm obs,w_p}$ is shown in the second and fourth row. \mycmt{$\pi_{\rm max}$ in the figures is for both observational and SHAM galaxies.} The deviation at $r_p$ around 20$\hmpc$ for the bulk sample ($0.2<z<0.43$) decreases while using a smaller $\pi_{\rm max}$. However, we see that the deviation at 5$\hmpc$ increases which might indicate systematics at smaller scales.}
    \label{wp lowz}
\end{figure}
\begin{figure}
    \centering
    \includegraphics[width=\linewidth]{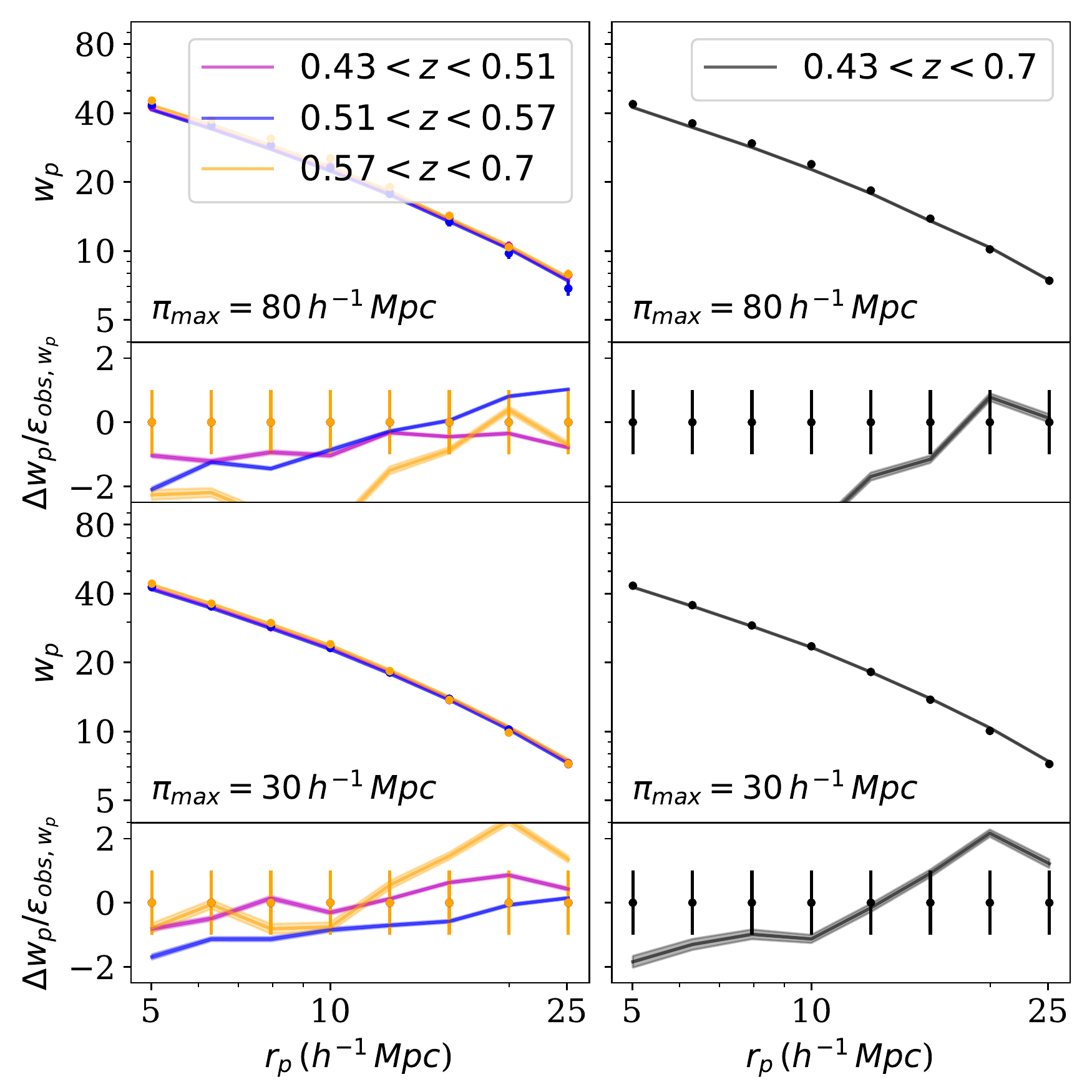}
    \caption{Same as \reffig{wp lowz}, but for CMASS LRGs. SHAM catalogues start to be more consistent with the observed $w_p$ with $\pi_{\rm max}=$ 30$\hmpc$ compared to $\pi_{\rm max}=$ 80$\hmpc$. Large deviations are found in the case of $\pi_{\rm max}=$ 80$\hmpc$. This is consistent with our findings from multipoles which indicate systematics at larger scales, e.g. $r > 18\hmpc$.}
    \label{wp cmass}
\end{figure}
\begin{figure}
    \centering
    \includegraphics[width=\linewidth]{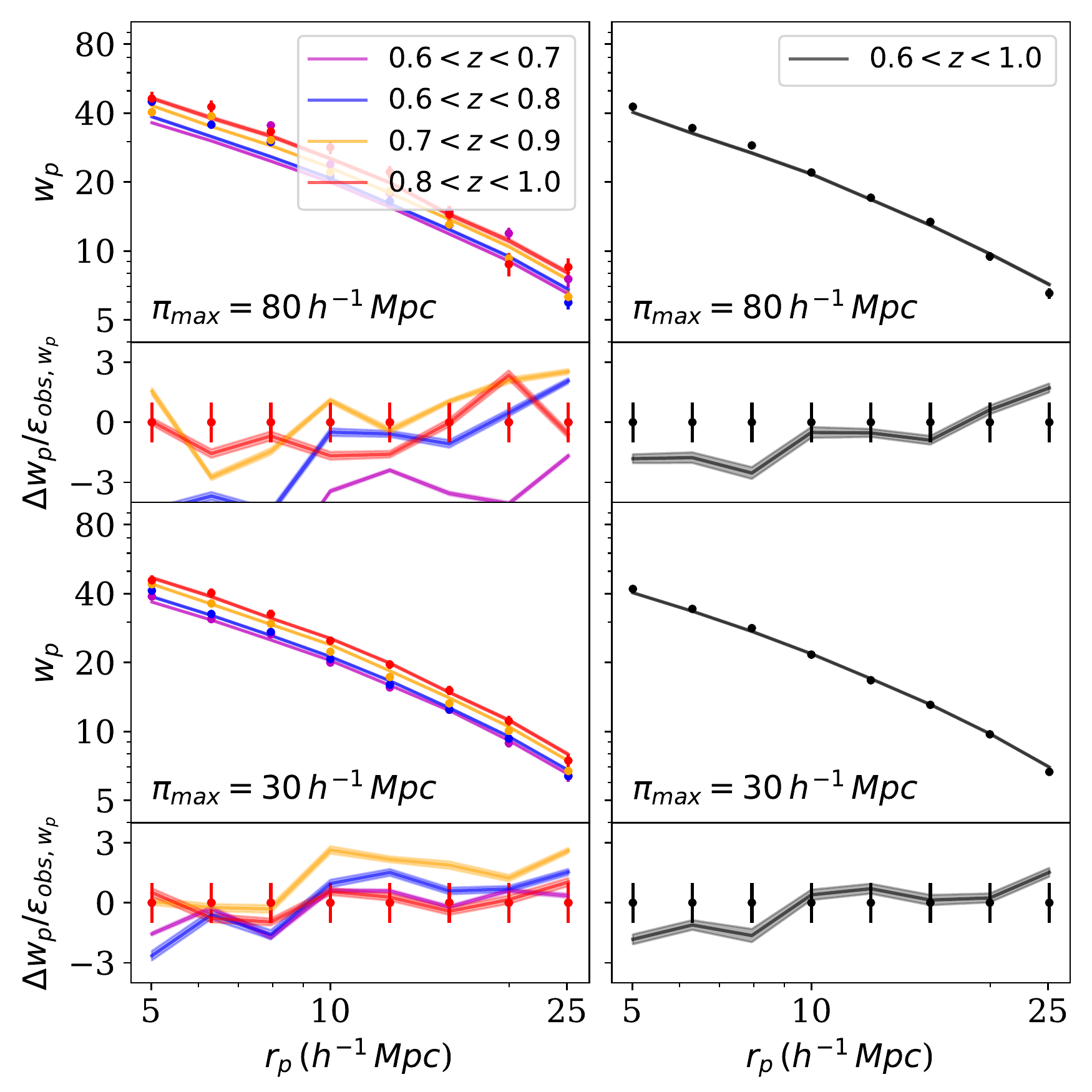}
    \caption{Same as \reffig{wp lowz}, but for eBOSS LRGs. Left figures (small redshift bins) show obvious deviations when using $\pi_{\rm max}=$ 80$\hmpc$, but the deviations are mitigated when using $\pi_{\rm max}=$ 30$\hmpc$. }
    \label{wp eboss}
\end{figure}
\subsection{Galaxy Incompleteness indicated by \texorpdfstring{$\sigma$}{sigma} and \texorpdfstring{$V_{\rm ceil}$}{Vceil}}
\label{sigma evolution}
Due to the high degeneracy between $\sigma$ and $v_{\rm ceil}$, the redshift evolutions of $\sigma$ and $V_{\rm ceil}$ are not obvious in the marginalized posterior of eBOSS (\reffig{eboss posterior}). But the 2D $\sigma$--$V_{\rm ceil}$ posterior contours show a clear trend that $\sigma$ or $V_{\rm ceil}$ are smaller at higher redshift. 

Thus, given a smaller $\sigma$, we keep more haloes with large $V_{\rm peak}$ that can be scattered to a certain range of $V_{\rm peak}^{\rm scat}$, leading to a smaller stellar mass incompleteness (see Section~\ref{result fitting}). A smaller $V_{\rm ceil}$ means to remove fewer haloes with the largest $V_{\rm peak}^{\rm scat}$, so the sample is more complete at massive-end. This agrees with the minimum magnitude truncation mentioned in Section~\ref{target selection}, since this threshold is expected to remove fewer galaxies at higher redshift \citep{Zhai_2017}. To demonestrate this, we plot the completeness--redshift relation in \reffig{comp eboss}. The completeness is defined as the ratio of the number of eBOSS LRG targets with a complete target selection including  \citep{prakash_sdss-iv_2016}, to that from a target selection without the $i$-band lower limit $i\ge 19.9$ (\refeq{iband}). Since there are no spectroscopic redshift measurements for objects excluded by the eBOSS target selection criteria, we match the eBOSS target catalogues with the DECaLS DR9 data\footnote{\url{https://www.legacysurvey.org/dr9/files}} \citep{decals} to make use of the DECaLS photometric redshift measurements \citep{rongpu} and 78.5 per cent of the eBOSS LRG candidates without the $i$-band lower limit are found in the DECaLS catalogue. 

For the complete samples, i.e., LOWZ and CMASS at $0.43<z<0.51$), we obtain much tighter constraints on $\sigma$ at different redshifts, as shown in \reftab{snapshot}. Since the $\sigma$ parameter in this study absorbs the incompleteness of stellar mass function, our measurements provide upper bounds of the intrinsic scatter between stellar mass and our halo mass proxy $V_{\rm peak}$, i.e. $\sigma_{\rm int} < 0.31$ at $0.2<z<0.33$, $\sigma_{\rm int} < 0.36$ at $0.33<z<0.43$, and $\sigma_{\rm int} < 0.46$ at $0.43<z<0.51$ with 95 percent confidence. We do not consider the measurement at $0.51<z<0.57$ to be robust as explained in Section~\ref{result fitting}.

\subsection{Consistency Check with \texorpdfstring{$w_p$}{wp}}
\label{wp consistency}
In principle, when the 2PCF multipoles of a SHAM catalogue agree well with those of the corresponding observational data, the $w_p$ of the two catalogues should also be consistent. To investigate the impacts of potential systematic effects that are usually observed on large scales \citep{ashley2017}, we measure the $w_p$ with two integral constraints for both the best-fitting SHAM catalogues and observations. At first, we use $w_p$ with $\pi_{\rm max} = 80\hmpc$ to obtain a good SNR and avoid the noise-dominated large scales \citep{Mohammad_2020}. This $\pi_{\rm max}$ value is a common choice for HOD studies \citep[e.g.,][]{ELGHOD2020}. In this case, the $w_p$ of the SHAM for LOWZ samples agrees with the observed $w_p$, while for CMASS and eBOSS samples, the $w_p$ of SHAM catalogues are generally underestimated compared to those of the observations. We then reduce $\pi_{\rm max}$ to $30\hmpc$, and the $w_p$ measured from the SHAM and observational catalogues become consistent, as shown in Figures~\ref{wp lowz}--\ref{wp eboss}. \mycmt{Note that $\pi_{\rm max}$ values are for both observational and SHAM $w_p$.}

Potential uncorrected systematics that affect the 2PCF monopole on $r \gtrsim 80\hmpc$ can explain the inconsistency in $w_p$ with $\pi_{\rm max}=80\hmpc$. The discrepancy on the large scales of the 2PCF monopole is observed when comparing observations with the galaxy mocks like \textsc{Patchy} \citep{kitaura_clustering_2016} and \textsc{EZmock} \citep{Zhao2020}. Meanwhile, \citet{huterer_calibration_2013} explain that uncorrected photometric systematics can significantly bias clustering on large scales.

\begin{figure}
    \centering
    \includegraphics[width=\linewidth,scale=0.42]{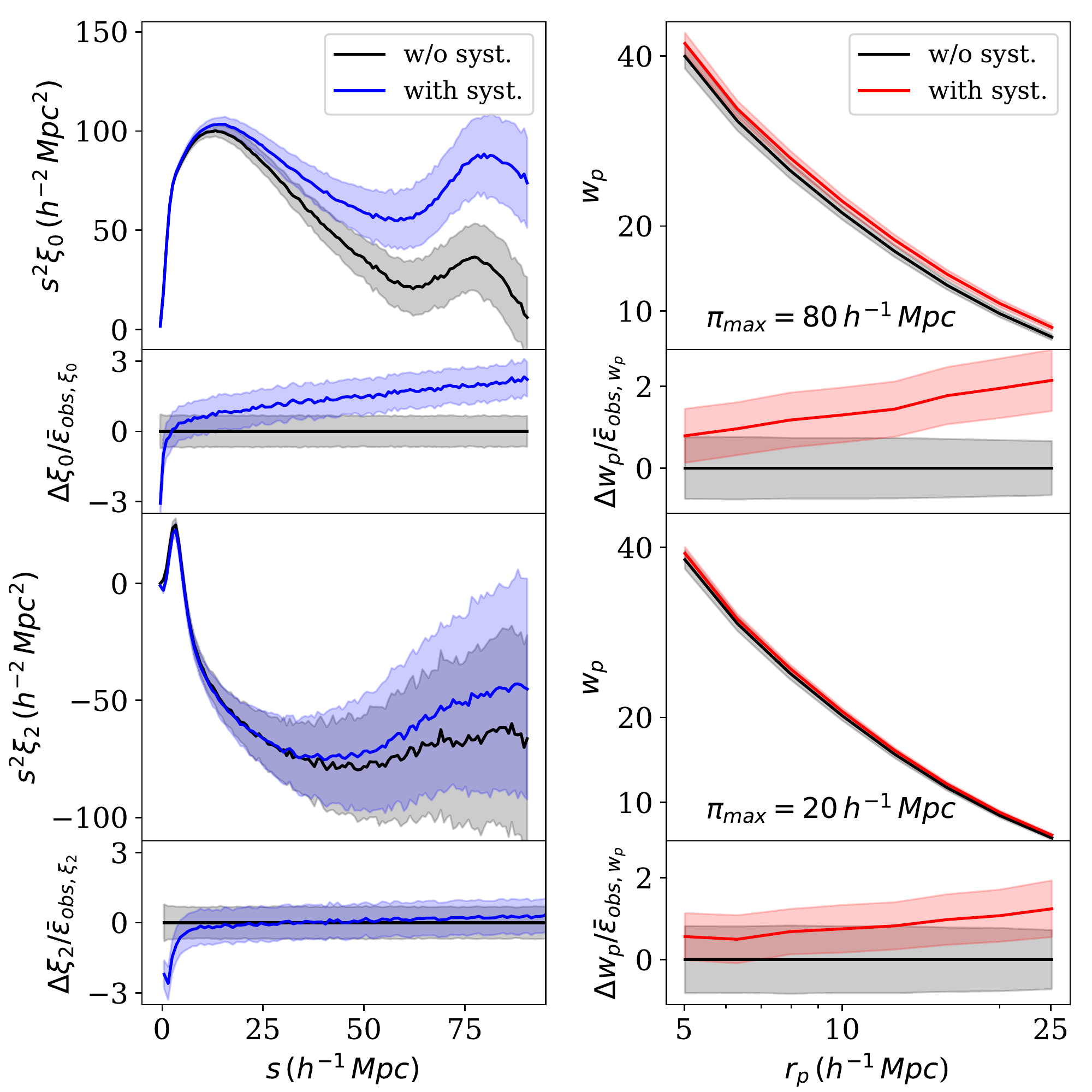}
    \caption{The effect of a biased 2PCF monopole on $w_p$. The first column represents the 2PCF multipoles of \textsc{EZmock} mocks without systematics (black lines with shades) and \textsc{EZmock} mocks with systematics (blue lines with shades). The second column is the $w_p$ comparison between \textsc{EZmock} mocks without systematics (black lines with shades) and \textsc{EZmock} mocks with systematics (red lines with shades). The first row is $w_p$ with $\pi$ integrating to 80$\hmpc$ and the third row is $\pi$ integrating to 20$\hmpc$. The second and fourth row for both columns are the difference between their clustering normalized by the quadrature sum of the standard deviations for \textsc{EZmock} mocks with and without systematics. The $w_p$ difference is suppressed when $\pi_{\rm max}$ equals a smaller value. }
    \label{EZmock mps}
\end{figure}

To directly demonstrate the consequences of uncorrected systematics on $w_p$ with different $\pi_{\rm max}$, we compare the 2PCF multiples and projected 2PCFs of \textsc{EZmock} mocks without and with observational systematics, respectively. \mycmt{\textsc{EZmock} mocks with observational systematics are mocks that mimic the data, including all the systematics mentioned in Section~\ref{galaxy weights}. We do not apply weight corrections, to investigate the potential biases due to systematics.} The clustering difference is normalized by the quadrature sum of the standard deviations for \textsc{EZmock} mocks with and without systematics. As is shown in \reffig{EZmock mps}, the observational systematics is responsible for the monopole difference, especially at scales larger than $50\hmpc$. Consequently, the bias presented in $w_p$ with $\pi_{\rm max}=80\hmpc$ is much larger than that of $w_p$ with $\pi_{\rm max}=20\hmpc$. The result reveals the importance of properly choosing $\pi_{\max}$ to avoid large-scale systematics as much as possible.  

\section{Conclusion}
\label{conclusion}
Subhalo Abundance matching is a powerful method of constructing galaxy catalogue based on high-resolution simulations. We propose a 3-parameter SHAM algorithm that is more general for different galaxy surveys. Besides the classical $\sigma$, we introduce $v_{\rm smear}$ that smears the peculiar velocities of SHAM galaxies to model the effect of the redshift uncertainties, and $V_{\rm ceil}$ that removes the most massive haloes/galaxies to account for the completeness of tracers in the massive end due to the target selection (e.g. luminosity threshold).

We construct SHAM catalogues that reproduce the clustering of BOSS/eBOSS LRGs on $[5, 25]\hmpc$ at $0.2<z<1.0$, based on the UNIT simulations. The 2PCF multipoles and projected 2PCFs of the SHAM catalogues are consistent with those of BOSS/eBOSS samples, including the entire LOWZ, CMASS and eBOSS samples, and 9 subsamples in different redshift slices. These results validate our new SHAM model. However, for some of the samples such as CMASS at $0.43<z<0.7$ and eBOSS LRG $0.7<z<0.9$, the best-fitting reduced $\chi^2$ values for the 2PCF multipoles can be larger than 2. This may be due to unknown observational systematics. For the bulk CMASS samples, the inhomogeneity in completeness at different redshift bins can also contribute to the disagreement. 

Since the galaxies at $z<0.6$ are supposed to be complete \citep{reid_sdss-iii_2016}, a SHAM model without $V_{\rm ceil}$ should also work for them. This is confirmed by the fitting results except for the CMASS sample at $0.51<z<0.57$, which is rejected at more than $4\sigma$. Therefore, the BOSS LRGs should have already been incomplete from $z\approx 0.5$. This finding is especially important for the reconstruction of the stellar mass function.

Using our SHAM method, the 2PCF bias brought by redshift uncertainties can be quantified through the $v_{\rm smear}$ parameter. Meanwhile, we estimate the redshift uncertainty statistically with the redshift difference $\Delta v$ obtained from repeat observations. There are two estimators of the statistical redshift uncertainties: (1) the best-fitting Gaussian dispersions $\sigma_{\Delta v}$ for the histogram of $\Delta v$; (2) the standard deviations of $\Delta v$ as $\hat{\sigma}_{\Delta v}$. As expected, both of them increase monotonically with the effective redshift due to broader absorption lines at higher redshifts. Both $\sigma_{\Delta v}$ and $\hat{\sigma}_{\Delta v}$ agree with our CMASS $v_{\rm smear}$ measurements. For eBOSS samples, $v_{\rm smear}$ agrees with the $\hat{\sigma}_{\Delta v}$ measurements while it is in tension with $\sigma_{\Delta v}$. This can be explained by the larger outlier fraction of eBOSS histograms. In short, the standard deviations of the redshift differences are in agreement with $v_{\rm smear}$ for both the CMASS and eBOSS LRG samples.

However, for the BOSS LOWZ sample, there are significant disagreements between $v_{\rm smear}$ values and the redshift uncertainties measured from repetitively observed targets. We have looked into some potential sources of systematic biases including the robustness of subhaloes from the UNIT simulations, the SDSS-III spectroscopic pipeline upgrade and the representativeness of the repeat samples. But none of them can lead to the observed discrepancy. Despite the fact that we have similar satellite fraction for LOWZ and CMASS, it is still possible that LOWZ sample is composed of a different subclass of LRGs that has different satellite properties compared to LRGs from CMASS and eBOSS. To validate this, we need a more detailed SHAM model for LOWZ samples. We leave relevant studies to a future work.

We observe the redshift evolution of eBOSS $\sigma$--$V_{\rm ceil}$ posteriors, indicating the sample is more complete at higher redshift bins. With the photometric redshift from DECaLS, we confirm the target selection criteria with a minimum i-band magnitude cut leads to the completeness evolution. Since the scatter parameter $\sigma$ in our study include both the intrinsic scatter $\sigma_{\rm int}$ and the incompleteness of galaxy samples, our 2-parameter SHAM measurements provide the constraints on $\sigma_{\rm int}$ as $\sigma_{\rm int}<0.31$ for LOWZ at $0.2<z<0.33$, $\sigma_{\rm int}<0.36$ for LOWZ at $0.33<z<0.43$, and $\sigma_{\rm int}<0.46$ for CMASS at $0.43<z<0.51$.

The projected 2PCFs of the best-fitting SHAM catalogues are consistent with those of the observations when the integral limit $\pi_{\rm max}$ is as small as $30\hmpc$. However, $w_p$ of SHAM catalogues deviates from the observations when $\pi_{\rm max} = 80\hmpc$. This can be explained by potential uncorrected systematics that mainly affect large scales. A test based on \textsc{EZmock} mocks with and without systematics also proves that $w_p$ evaluated with a $\pi_{\rm max}$ of around $30\hmpc$ is less sensitive to systematic effects, compared to $w_p$ with $\pi_{\rm max}=80\hmpc$. The result also supports the choice of a smaller $\pi_{\rm max}$ when using $w_p$ for SHAM and HOD studies \citep[e.g.,][]{Shadab2020}.

To conclude, our 3-parameter SHAM model works well for LRGs at a wide range of redshift, though there are a few exceptions that may be due to uncorrected observational systematics. Our algorithm is also effective in estimating uncertainties of redshift measurements, and detection of sample incompleteness for most of the SDSS samples. We are going to further improve the SHAM model by accounting for more properties, such as the satellite fraction. In the meantime, we also plan to extend our the studies to ELGs. Because their star-forming processes are quenched in massive haloes \citep{kauffmann_environmental_2004,Dekel2006}, the parameter $V_{\rm ceil}$ is foreseen to be crucial. Later, we shall also construct multi-tracer SHAM models for different types of tracers with both auto- and cross-correlations for cosmological analysis.

\section*{Data Availability}
The LOWZ and CMASS clustering and random catalogues, \textsc{Patchy} mocks are all from SDSS data release 12. For eBOSS, the PIP+ANG weighted clusterings can be provided by FM. The corresponding \textsc{EZmock} mocks can be obtained with request to CZ. The \textsc{spAll} catalogue used for the redshift difference measurements are publicly available and the code of the redshift uncertainty measurement can be obtained upon the request to JB. 

\section*{Acknowledgement}
JY, CZ and GF acknowledge support from the SNF 200020\_175751 “Cosmology with 3D Maps of the Universe” research grant. GR acknowledges support from the National Research Foundation of Korea (NRF) through Grant No. 2020R1A2C1005655 funded by the Korean Ministry of Education, Science and Technology (MoEST).

Funding for the Sloan Digital Sky Survey IV has been provided by the Alfred P. Sloan Foundation, the U.S. Department of Energy Office of Science, and the Participating Institutions. SDSS-IV acknowledges support and resources from the Center for High-Performance Computing at the University of Utah. The SDSS web site is \url{www.sdss.org}.

SDSS-IV is managed by the Astrophysical Research Consortium for the Participating Institutions of the SDSS Collaboration including the 
Brazilian Participation Group, the Carnegie Institution for Science,
Carnegie Mellon University, the Chilean Participation Group,
the French Participation Group, Harvard-Smithsonian Center for Astrophysics, 
Instituto de Astrof\'isica de Canarias, The Johns Hopkins University,
Kavli Institute for the Physics and Mathematics of the Universe (IPMU) / University of Tokyo,
the Korean Participation Group, Lawrence Berkeley National Laboratory, 
Leibniz Institut f\"ur Astrophysik Potsdam (AIP),  
Max-Planck-Institut f\"ur Astronomie (MPIA Heidelberg), 
Max-Planck-Institut f\"ur Astrophysik (MPA Garching), 
Max-Planck-Institut f\"ur Extraterrestrische Physik (MPE), 
National Astronomical Observatories of China, New Mexico State University, 
New York University, University of Notre Dame, 
Observat\'ario Nacional / MCTI, The Ohio State University, 
Pennsylvania State University, Shanghai Astronomical Observatory, 
United Kingdom Participation Group,
Universidad Nacional Aut\'onoma de M\'exico, University of Arizona, 
University of Colorado Boulder, University of Oxford, University of Portsmouth, 
University of Utah, University of Virginia, University of Washington, University of Wisconsin, 
Vanderbilt University, and Yale University.

\bibliographystyle{mnras}
\bibliography{reference} 


\appendix
\section{The effect of fibre collision}
\label{10-25 no bias}
\begin{figure}
    \centering
    \includegraphics[width=\linewidth]{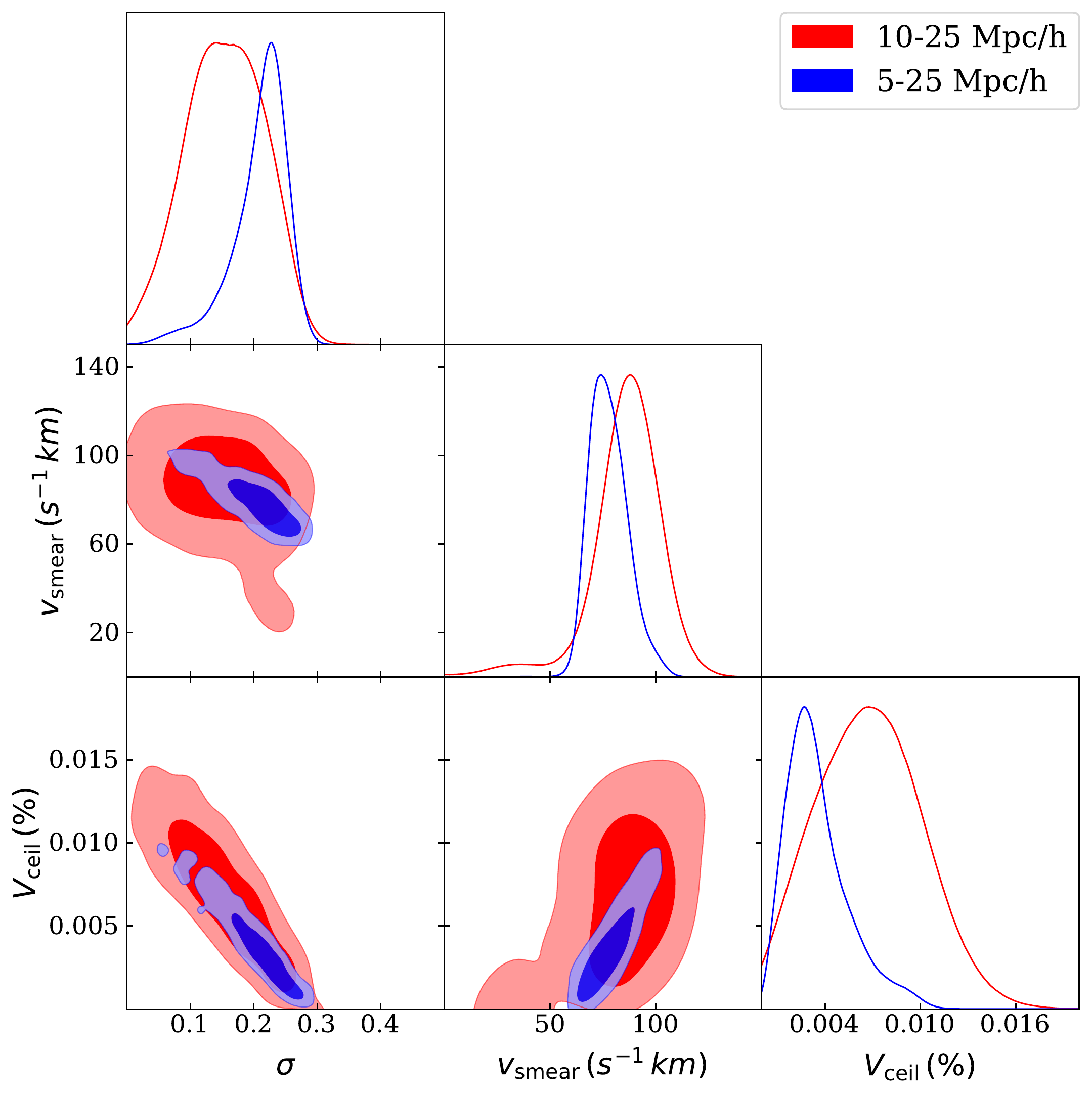}
    \caption{The posteriors of SHAM parameters for LOWZ at $0.2<z<0.43$, obtained with the fitting of monopole and quadropole at ranges on [10, 25]$\hmpc$ (red) and [5, 25]$\hmpc$ (blue).} 
    \label{10-25mpc/h}
\end{figure}

\mycmt{To avoid potential biases of the clustering measurements due to uncorrected fibre collision effects with nearest-neighbour close-pair weights ($w_{\rm CP}$),
we conduct a new SHAM fitting on the monopole and quadropole in [10, 25]$\hmpc$, which are not significantly affected by fibre collision effects \citep{Guo2012, Rodriguez-Torres2016},
for LOWZ at $0.2<z<0.43$. The resulting posterior distributions of parameters are shown in \reffig{10-25mpc/h}. As demonstrated in the figure, the constraints are consistent with those from our fiducial fitting range ([5, 25]$\hmpc$) at 1$\sigma$ level. It thus suggests that our SHAM fitting results are robust.
} 

\section{Blue Samples in LOWZ}
\label{blue-red ratio}
\begin{figure}
    \centering
    \includegraphics[width=\linewidth]{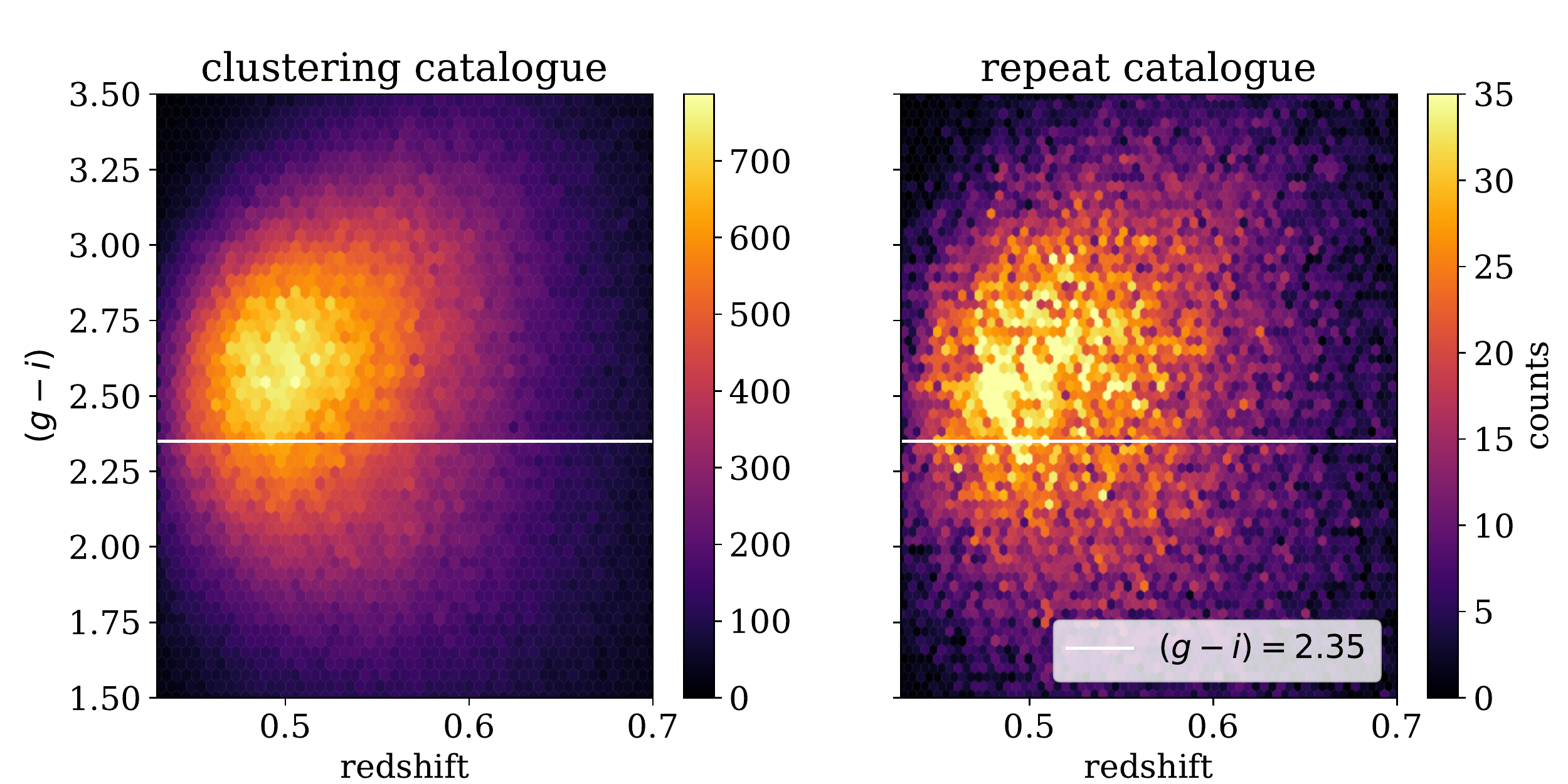}
    \caption{The colour--redshift diagram for CMASS at $0.43<z<0.7$ and the blue galaxy threshold (white dashed lines). \textit{Left:} the colour distribution of galaxies from the clustering catalogue. \textit{Right:} the colour distribution of galaxies from the repeat observation. } 
    \label{colourcut cmass}
\end{figure}

\begin{figure}
    \centering
    \includegraphics[width=\linewidth]{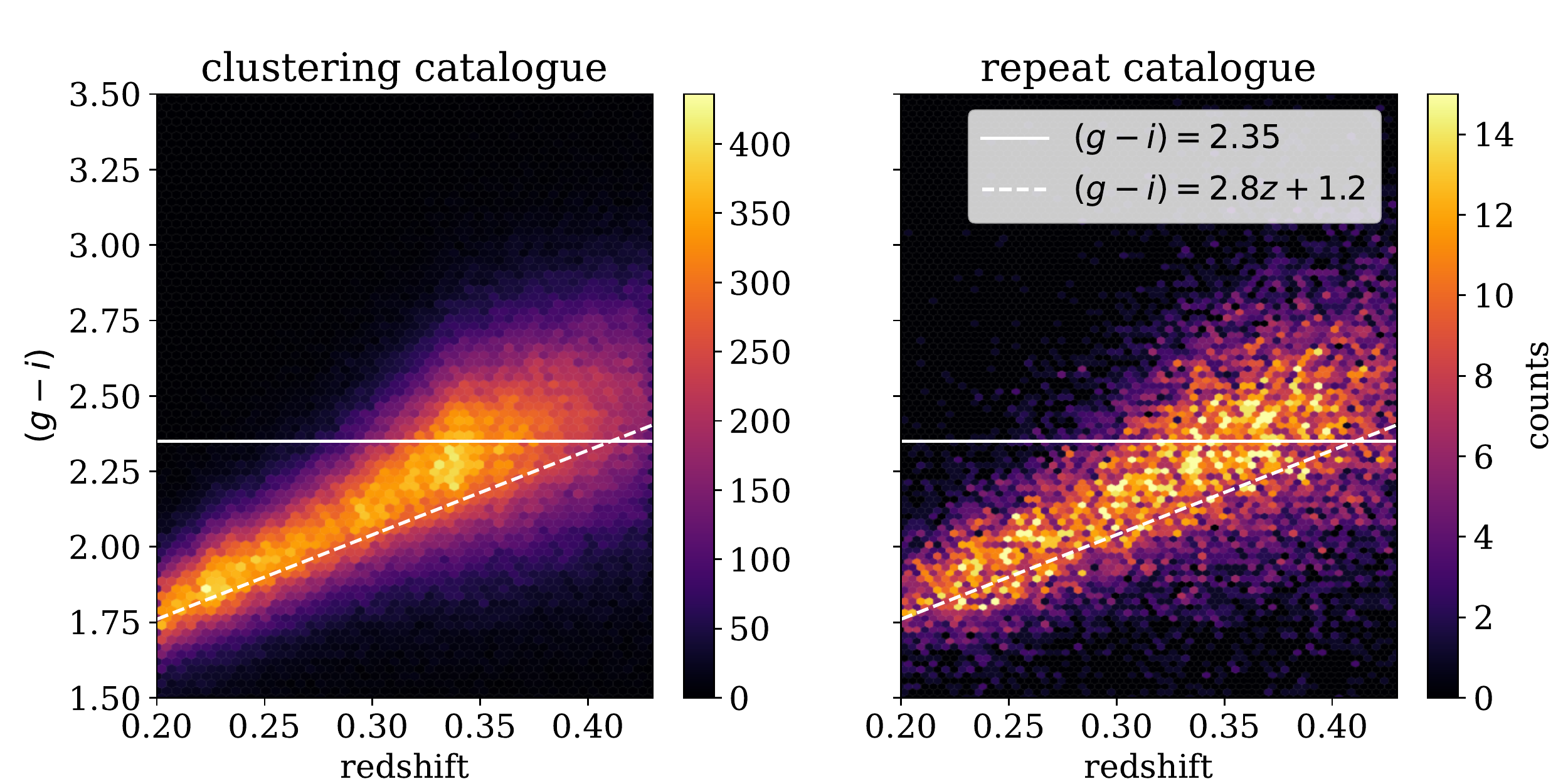}
    \caption{The same as \reffig{colourcut cmass} for LOWZ with two cuts. The first cut is $(g-i)=2.35$ (the white solid line) and the second one is $(g-i) = 2.8z+1.2$ (the white dashed line).} 
    \label{colourcut lowz}
\end{figure}

In order to indirectly prove the representative of the repeat observation, we choose to study the fraction of blue galaxies, i.e., $f_{\rm blue}$ in the repeat measurement catalogue and the galaxy clustering catalogue. The differentiation of the blue and red samples is achieved by applying an \textit{ad hoc} criterion on the colour--redshift diagram. For example, with a constant colour cut $(g-i)=2.35$, \citet{favole_building_2016} selects the blue tail of CMASS samples and find out that they have different clustering properties from the red samples. The magnitudes for colour calculation are the Composite Model Magnitudes \citep[i.e., \textsc{CMODELMAG};][]{reid_sdss-iii_2016} from the \textsc{spAll} catalogue. We start with using the same cut on both clustering catalogue and repeated samples for LOWZ and CMASS (\reffig{colourcut cmass} and \reffig{colourcut lowz}). The results in \reftab{blue ratios cut 1} shows that $f_{\rm blue}$ in clustering samples are quite close to $f_{\rm blue}$ of the repeat samples. Despite of the minor difference, for LOWZ, $f_{\rm blue}$ of the clustering catalogue is always larger than $f_{\rm blue}$ of the repeated samples, while for CMASS it is the opposite. Additionally, there is no significant difference in the $\hat{\sigma}_{\Delta v}$ of blue galaxies and red galaxies that can lead to the disagreement between LOWZ $v_{\rm smear}$ and LOWZ $\Delta v$. 


As shown in \reffig{colourcut lowz}, a constant cut cannot help to select the blue tail of LOWZ. A cut varies with the redshift is a better choice. So we apply $(g-i) = 2.8z+1.2$, and find $f_{\rm blue}$ drops to the same level as CMASS with a constant cut. The clustering catalogue still has $f_{\rm blue}$ larger than that of the repeat observations and we have small difference between the red and blue $\hat{\sigma}_{\Delta v}$ just like our findings in the results of the constant cut. The results can be found in \reftab{blue ratios cut 2}. So it proves that the redshift uncertainties measured based on repeat samples should be representative for the clustering sample.
\begin{table}
\centering
{\begin{tabular}[c]{|c|c|c|c|c|}
\hline
{ z range} & {$f_{\rm blue}$}   & {$f_{\rm blue}$} & {red $\hat{\sigma}_{\Delta v}$}& {blue $\hat{\sigma}_{\Delta v}$} \\
{} & {clst. (\%)}   & {repeat (\%)} & {$({\rm s}^{-1}\,{\rm km})$}& {$({\rm s}^{-1}\,{\rm km})$} \\
\hline
\hline
{$0.2<z<0.33$}  &{88.9} &{87.3} &{29.3$\pm$3.4} &{26.0$\pm$2.5} \\
\hline
{$0.33<z<0.43$} &{46.1} &{44.6} &{33.1$\pm$1.7} &{33.1$\pm$3.6} \\
\hline
{$0.2<z<0.43$}  &{67.3} &{64.6} &{32.5$\pm$1.5} &{28.8$\pm$2.2} \\
\hline
\hline
{$0.43<z<0.51$} &{37.3} &{37.5} &{46$\pm$2.7} &{48.3$\pm$4.7}\\
\hline
{$0.51<z<0.57$} &{36.3}&{37.4}  &{52.8$\pm$3.4} &{52.6$\pm$3.0}\\
\hline
{$0.57<z<0.7$} &{37.0}  &{38.8} &{61.0$\pm$3.6} &{60.7$\pm$4.8}\\
\hline
{$0.43<z<0.7$}  &{36.9}&{38.0} &{54.0$\pm$1.9} &{54.6$\pm$2.8}\\
\hline
\end{tabular}}
\caption{The $f_{\rm blue}$ for the clustering catalogue and the repeat observation, and the $\hat{\sigma}_{\Delta v}$ of the blue and the red galaxies for LOWZ and CMASS samples at different redshift bins using the constant cut $(g-i)=2.35$. The $f_{\rm blue}$ values of LOWZ clustering catalogue are larger than those of the repeat catalogue, while CMASS samples shows the opposite relation. The difference between the red and blue $\hat{\sigma}_{\Delta v}$ is not significant for LOWZ and CMASS.}
\label{blue ratios cut 1} 
\end{table}
\begin{table}
\centering
{\begin{tabular}[c]{|c|c|c|c|c|}
\hline
{ z range} & {$f_{\rm blue}$}   & {$f_{\rm blue}$} & {red $\hat{\sigma}_{\Delta v}$}& {blue $\hat{\sigma}_{\Delta v}$} \\
{} & {clst. (\%)}   & {repeat (\%)} & {$({\rm s}^{-1}\,{\rm km})$}& {$({\rm s}^{-1}\,{\rm km})$} \\

\hline
\hline
{$0.2<z<0.33$}  &{30.9} &{29.1} &{26.8$\pm$2.9} &{25.7$\pm$1.7} \\
\hline
{$0.33<z<0.43$} &{30.5} &{30.2} &{32.7$\pm$1.4} &{34.1$\pm$5.1} \\
\hline
{$0.2<z<0.43$}  &{30.7} &{29.7} &{30.0$\pm$1.5} &{30.5$\pm$3.2} \\
\hline
\end{tabular}}
\caption{The same as \reftab{blue ratios cut 1} for LOWZ using the redshift cut $(g-i)=2.8z+1.2$. The $f_{\rm blue}$s of clustering are still larger than those of the repeat catalogue and the difference between the red and blue $\hat{\sigma}_{\Delta v}$ is also small.}
\label{blue ratios cut 2} 
\end{table}

\section{Posteriors of BOSS and eBOSS SHAM}
\label{appendix posteriors}
\begin{figure}
    \centering
    \includegraphics[width=\linewidth]{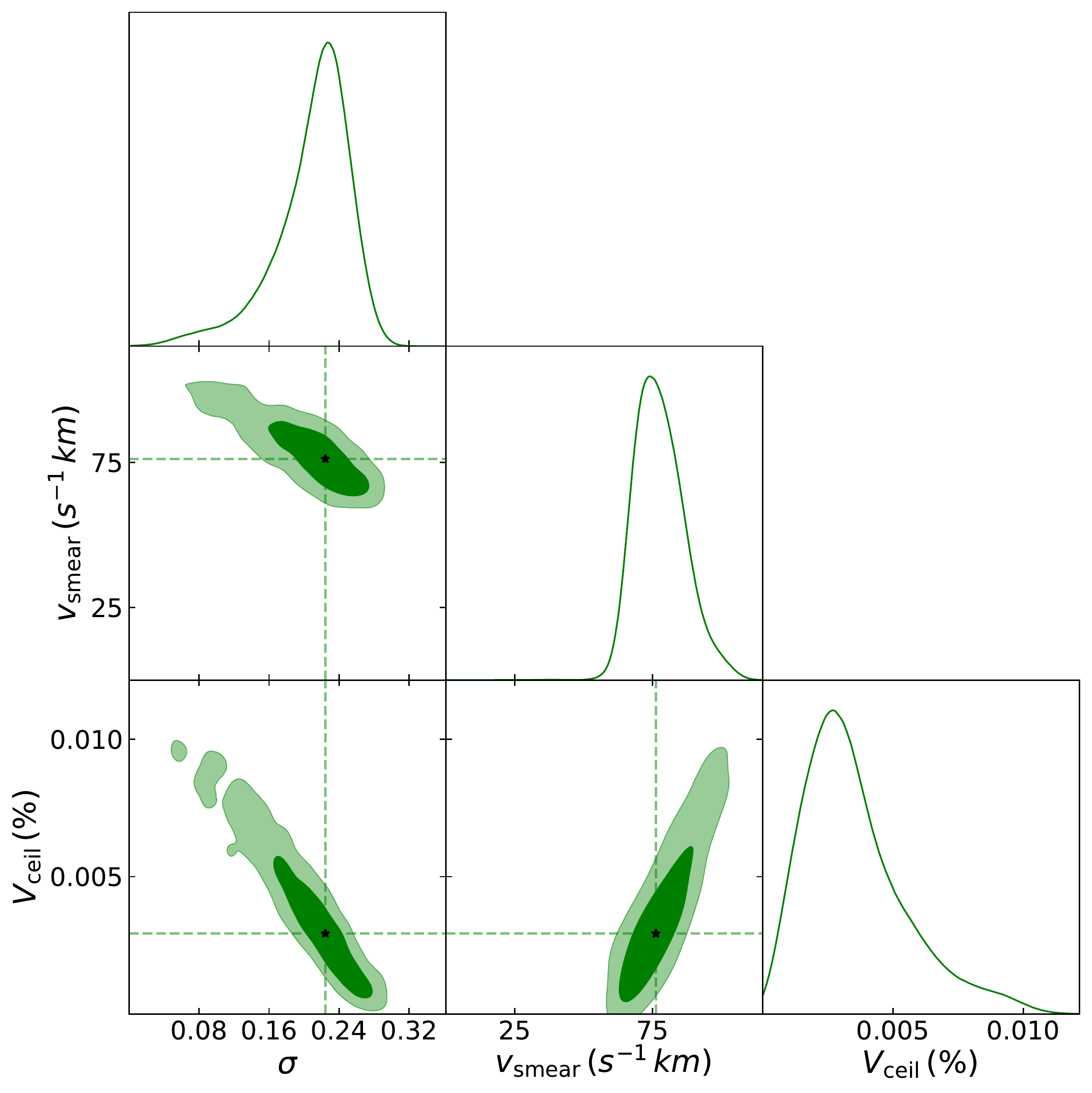}
    \caption{The posteriors of SHAM parameters for LOWZ samples at $0.2<z<0.43$. The black dots in the contours indicate the position of the best-fitting parameters.} 
    \label{lowz bulk posterior}
\end{figure}

\begin{figure}
    \centering
    \includegraphics[width=\linewidth]{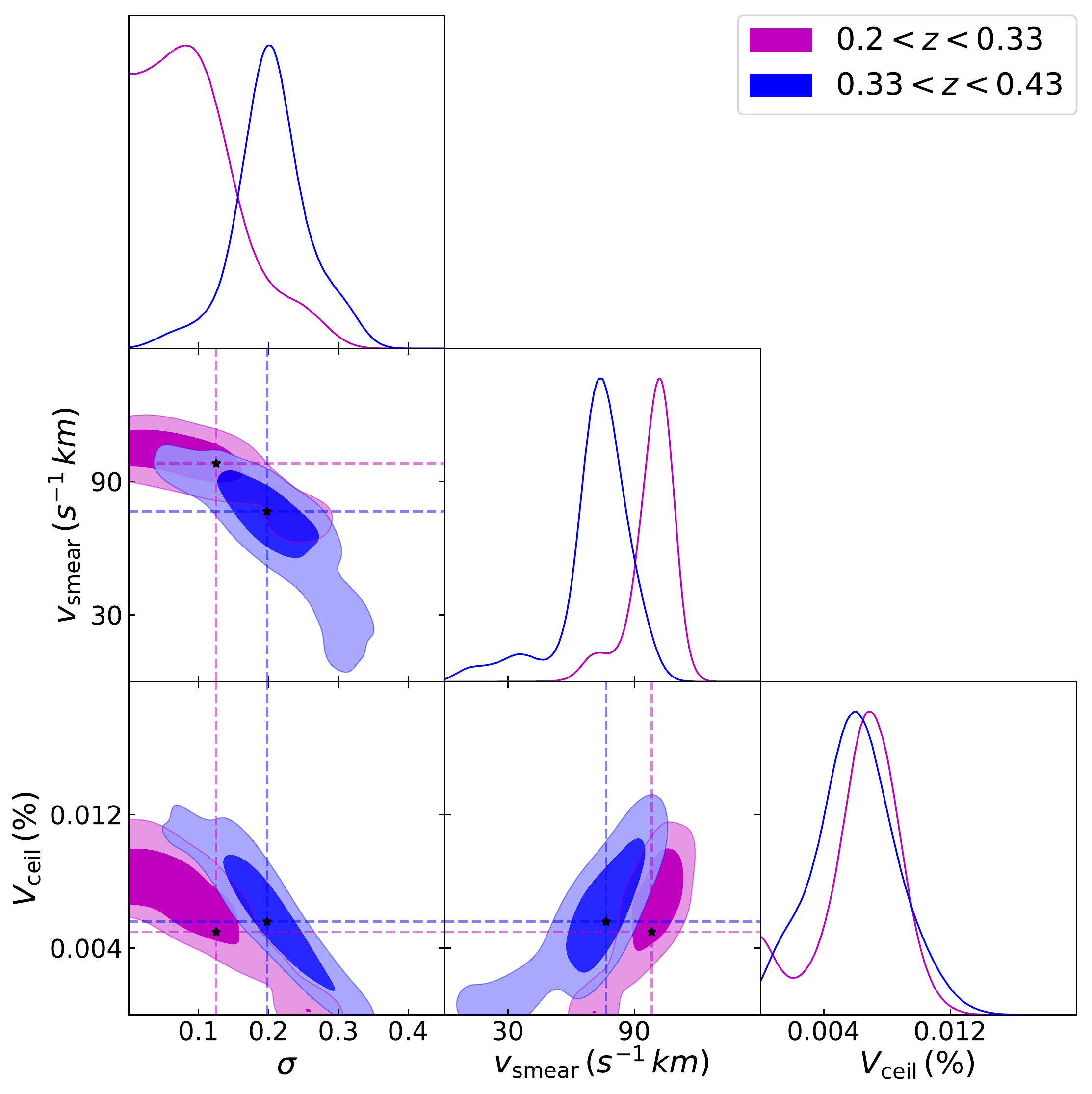}
    \caption{The same as \reffig{lowz bulk posterior} for LOWZ SHAM at $0.2<z<0.33$ (magenta) and $0.33<z<0.43$ (blue).}  
    \label{lowz posterior}
\end{figure}

\begin{figure}
    \centering
    \includegraphics[width=\linewidth]{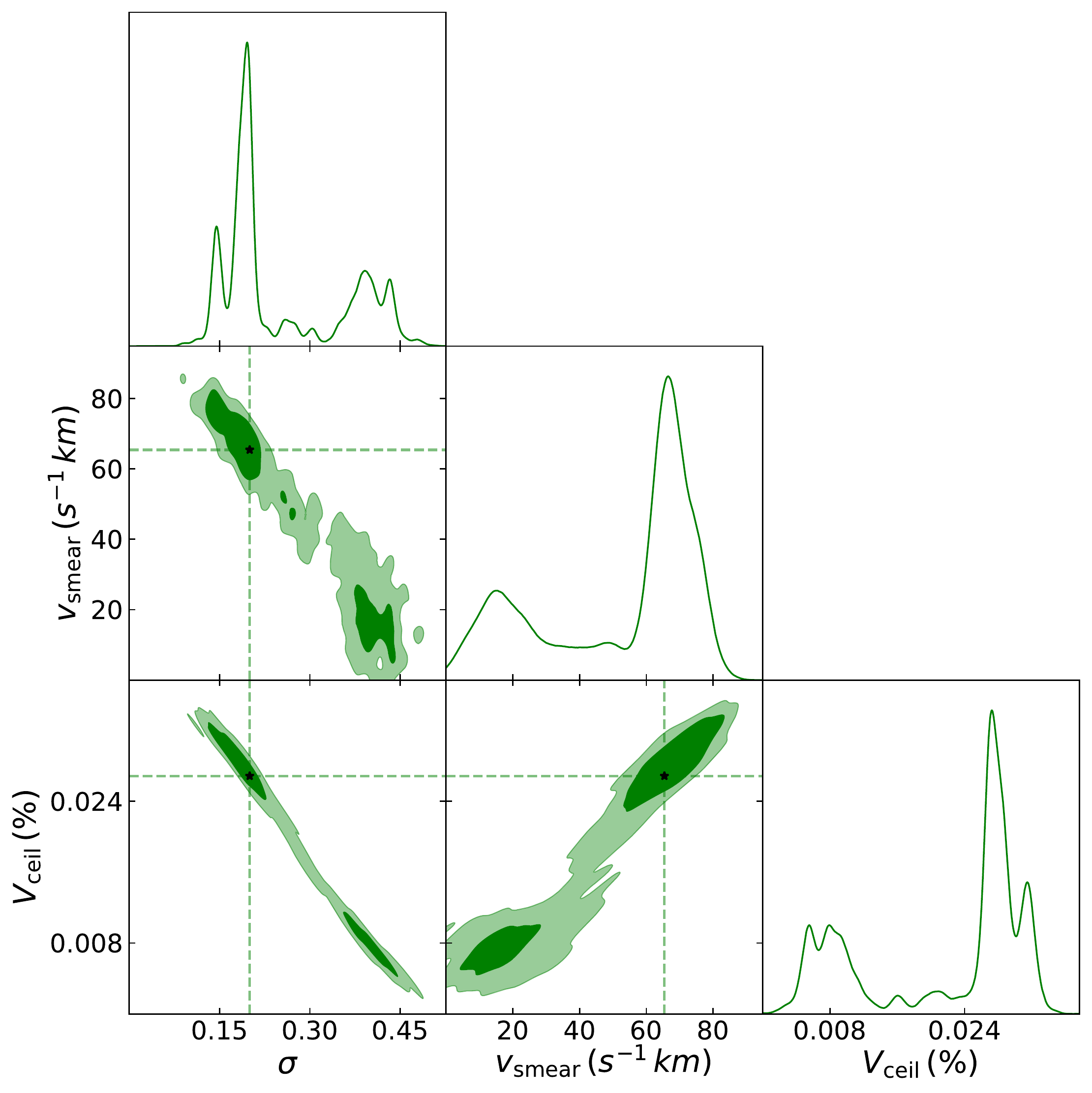}
    \caption{The same as \reffig{lowz bulk posterior} for CMASS SHAM at $0.43<z<0.7$.}  
    \label{cmass bulk posterior}
\end{figure}

\begin{figure}
    \centering
    \includegraphics[width=\linewidth]{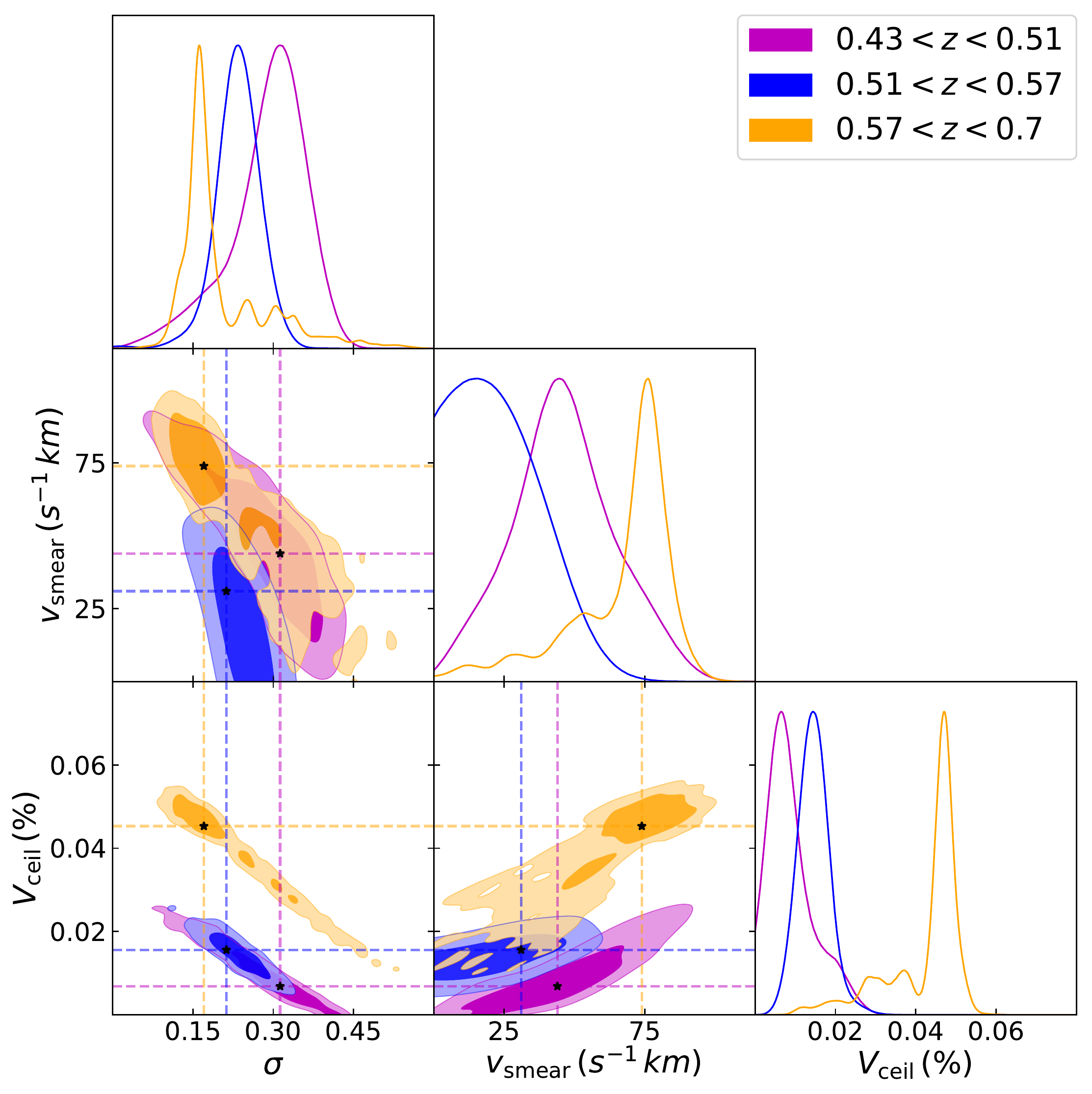}
    \caption{The same as \reffig{lowz bulk posterior} for LOWZ SHAM at $0.43<z<0.51$ (magenta), $0.51<z<0.57$ (blue) and $0.57<z<0.7$ (orange).}
    \label{cmass posterior}
\end{figure}

\begin{figure}
    \centering
    \includegraphics[width=\linewidth]{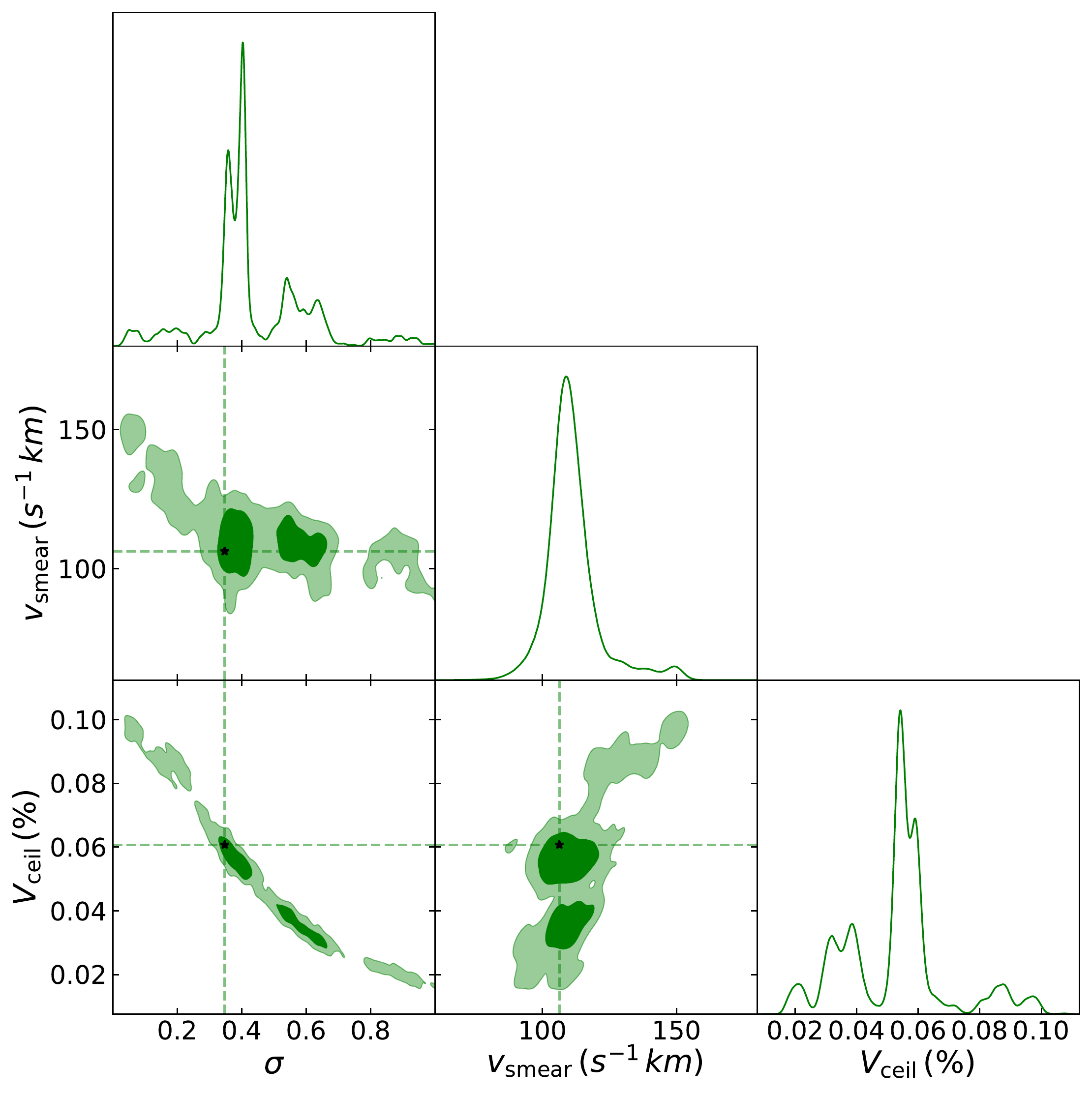}
    \caption{The same as \reffig{lowz bulk posterior} for eBOSS LRG SHAM at $0.6<z<1.0$.}  
    \label{eboss bulk posterior}
\end{figure}

\begin{figure}
    \centering
    \includegraphics[width=\linewidth]{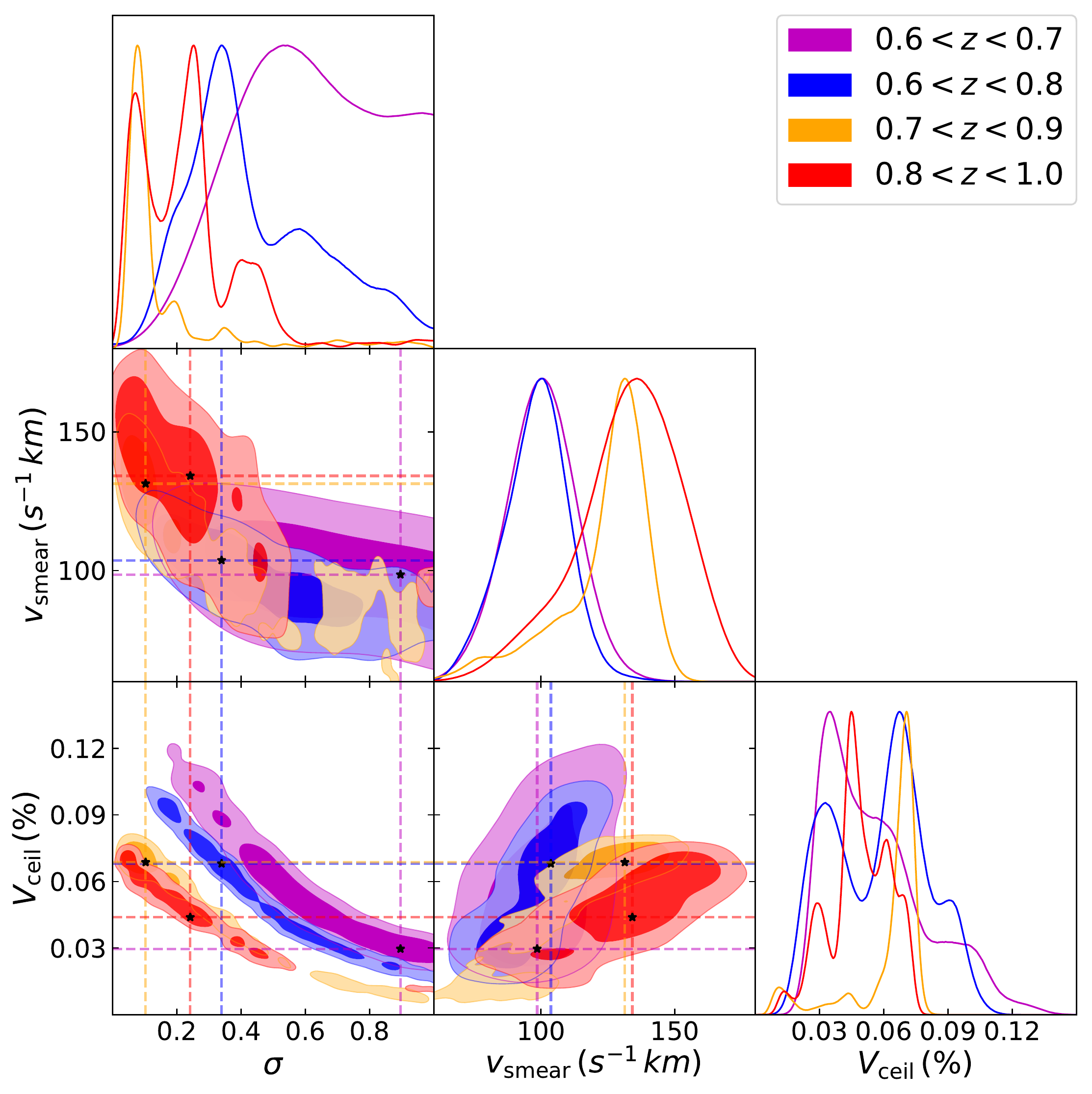}
    \caption{The same as \reffig{lowz bulk posterior} for eBOSS LRG SHAM at $0.6<z<0.7$ (magenta), $0.6<z<0.8$ (blue), $0.7<z<0.9$ (orange) and $0.8<z<1.0$(red). The evolution of the $\sigma$--$V_{\rm ceil}$ contour indicates the incompleteness at higher redshift bins is lower than that of the lower redshift bins as explain in Section~\ref{sigma evolution}. }  
    \label{eboss posterior}
\end{figure}

The posteriors of all the 3-parameter SHAM results are presented in Figures~\ref{lowz bulk posterior}--\ref{eboss posterior}. The top panel of each column shows the marginalized posterior of the parameter written at the bottom. The remaining three panels are the 2D posterior contours. Parameters with the maximum likelihood determined by \textsc{pyMultinest}\footnote{\url{https://github.com/JohannesBuchner/PyMultiNest}} \citep{pymultinest} are marked in the figures. \mycmt{All the posteriors are from converged Monte-Carlo chains of \textsc{Multinest}. The multi-modal 1D posteriors of $\sigma$ and $V_{\rm ceil}$ may be due to the noisy data.} 

\begin{figure}
    \centering
    \includegraphics[width=\linewidth]{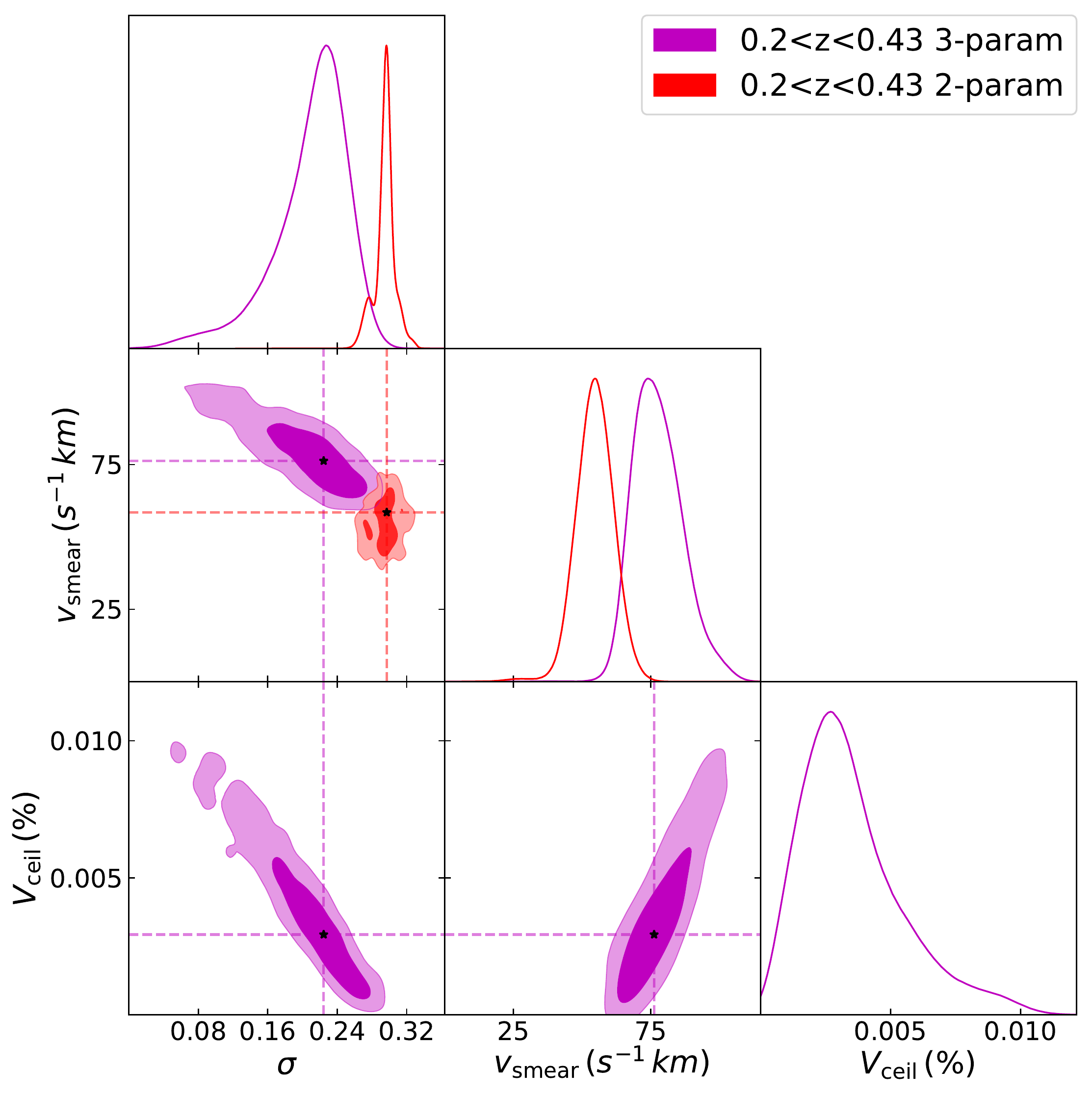}
    \caption{The posteriors of 3-parameter SHAM (magenta) compared with 2-parameter SHAM (red) for LOWZ samples at $0.2<z<0.43$.} 
    \label{2-param LOWZtot}
\end{figure}

\begin{figure}
    \centering
    \includegraphics[width=\linewidth]{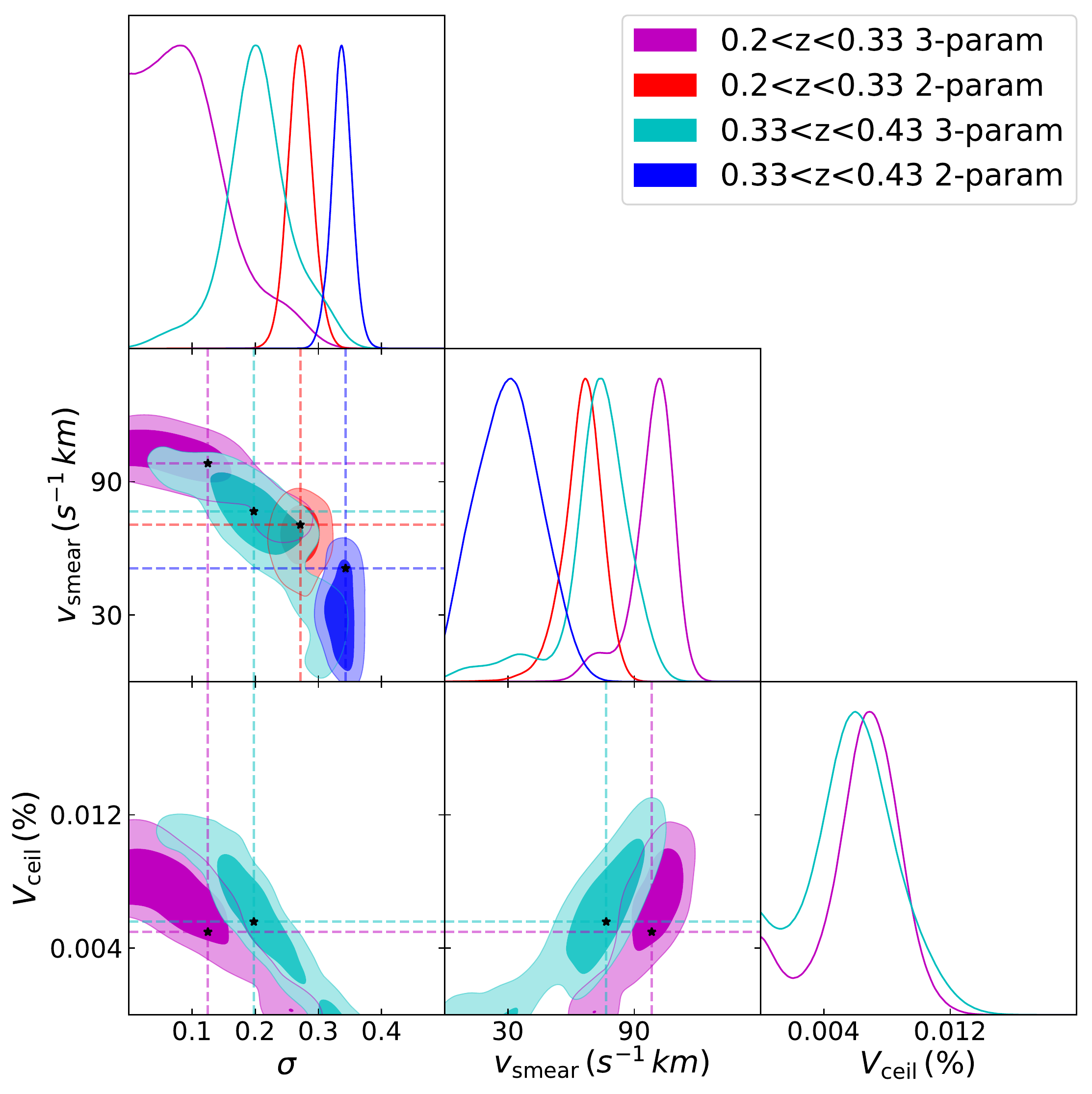}
    \caption{The same as \reffig{2-param LOWZtot} for LOWZ SHAM comparisons at $0.2<z<0.33$ (magenta, red) and $0.33<z<0.43$ (cyan, blue).}  
    \label{2-param LOWZ}
\end{figure}

\begin{figure}
    \centering
    \includegraphics[width=\linewidth]{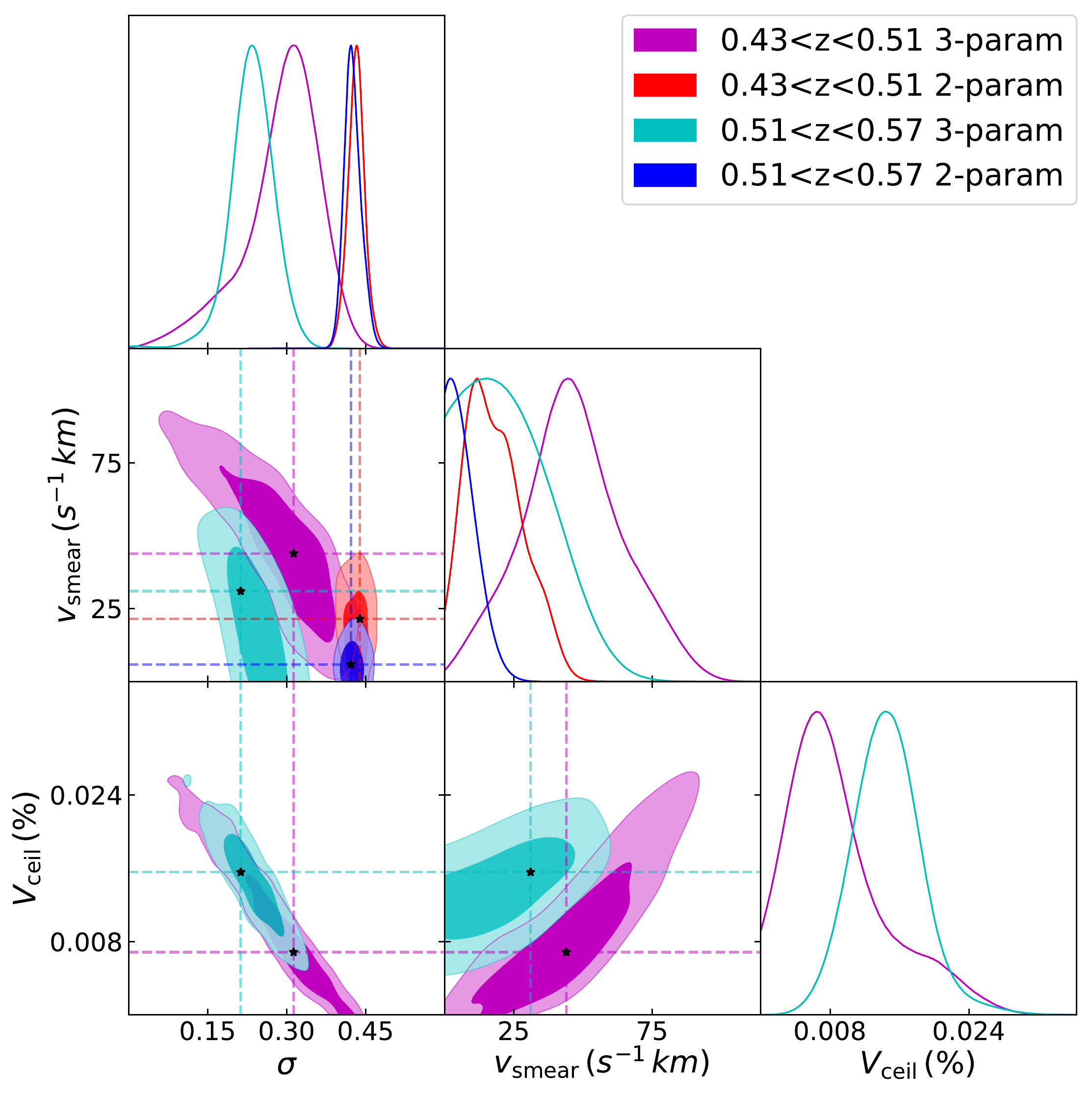}
    \caption{The same as \reffig{2-param LOWZtot} for CMASS SHAM comparisons at $0.43<z<0.51$ (magenta, red) and $0.51<z<0.57$ (cyan, blue).}
    \label{2-param CMASS}
\end{figure}

\mycmt{We also present the posteriors of the 2-parameter SHAM without $V_{\rm ceil}$ in Figures~\ref{2-param LOWZtot}--\ref{2-param CMASS}, comparing with those of the corresponding 3-parameter SHAM. We keep the posteriors of $V_{\rm ceil}$ for the convenience of visualizing the $V_{\rm ceil}$--$v_{\rm smear}$ degeneracy.}

\bsp	
\label{lastpage}
\end{document}